\documentclass[twocolumn]{aastex63}

\accepted{October 11, 2019}

\submitjournal{ApJ}

\shorttitle{Nucleosynthesis Constraints on the Energy Growth Timescale of a CCSN Explosion}
\shortauthors{Sawada et al.}

\usepackage{amsmath}
\usepackage{natbib}
\usepackage{threeparttable}
\usepackage{rotating}

\begin{document}

\title{Nucleosynthesis Constraints on the Energy Growth Timescale of a Core-Collapse Supernova Explosion}

\correspondingauthor{Ryo Sawada}
\author{Ryo Sawada}
\affil{Department of astronomy, Kyoto University, Kitashirakawa-Oiwake-cho, Sakyo-ku, Kyoto 606-8502, Japan}
\email{ryo@kusastro.kyoto-u.ac.jp}
\author{Keiichi Maeda}
\affil{Department of astronomy, Kyoto University, Kitashirakawa-Oiwake-cho, Sakyo-ku, Kyoto 606-8502, Japan}

\begin{abstract}
Details of the explosion mechanism of core-collapse supernovae (CCSNe) are not yet fully understood. 
There is an increasing number of numerical examples by ab-initio core-collapse simulations leading to an explosion. 
Most, if not all, of the ab-initio core-collapse simulations represent a `slow' explosion in which the observed explosion energy ($\sim 10^{51}$ ergs) is reached in a timescale of $\gtrsim1$ second.
It is, however, unclear whether such a slow explosion is consistent with observations.
In this work, by performing nuclear reaction network calculations for a range of the explosion timescale $t_{\rm grow}$, from the rapid to slow models, we aim at providing nucleosynthetic diagnostics on the explosion timescale. 
We employ one-dimensional hydrodynamic and nucleosynthesis simulations above the proto-neutron star core, by parameterizing the nature of the explosion mechanism by $t_{\rm grow}$. 
The results are then compared to various observational constraints; the masses of $^{56}$Ni derived for typical CCSNe, the masses of $^{57}$Ni and $^{44}$Ti observed for SN 1987A, and the abundance patterns observed in extremely metal-poor stars. 
We find that these observational constraints are consistent with the `rapid' explosion ($t_{\rm grow} \lesssim 250$ ms), and especially the best match is found for a nearly instantaneous explosion ($t_{\rm grow} \lesssim 50$ ms). 
Our finding places a strong constraint on the explosion mechanism; the slow mechanism ($t_{\rm grow} \gtrsim 1000$ ms) would not satisfy these constraints, and the ab-inito simulations will need to realize a rapid explosion.  
\end{abstract}
\section{Introduction} \label{ sec:intro }

Core-collapse supernovae (CCSNe) occur at the end of the lives of massive stars ($M_{\rm ZAMS} > 8 M_\odot$). 
As the core of the star gravitationally collapses to a neutron star or a black hole, a shock wave is triggered on a newly-formed compact remnant, leading to the SN explosion \citep{1934PNAS...20..254B}. 
However, the detailed nature of the explosion mechanism remains unclear. The most promising scenario is the delayed neutrino-driven explosion \citep{1985ApJ...295...14B}. 
While an SN explosion had not been reproduced by numerical simulations for a few decades (e.g., \citealt{2000ApJ...539L..33R}; \citealt{2005ApJ...629..922S}), the situation has changed recently. 
There are now an increasing number of numerical examples leading to an explosion, even with ab-initio simulations in which multi-dimensional hydrodynamics equations are coupled with a detailed treatment of neutrino transport (see e.g., \citealt{2012ARNPS..62..407J} and references therein). 
However, the ab-initio simulations have some limitation. 
Especially, it is unclear whether the nature of the explosion shown by these simulations is consistent with observations, because a multi-dimensional, especially three-dimensional, simulation with a detailed treatment of microphysics is computationally too expensive to allow long-term investigations.

Such a long-term simulation is required to address the consistency with observations, in particular, to predict nucleosynthetic yields of supernovae. 
Recently, thus, there are also a number of `artificial-explosion' simulations, that is, the readily calculable supernova simulations in which the neutrino transfer is solved but with the neutrino luminosity calibrated to match to some well-studied SNe (e.g., \citealt{2015ApJ...806..275P}; \citealt{2016ApJ...821...38S}). 
These studies, especially spherically symmetric models, can be calculated systematically for various progenitor models, from the onset of the explosion up to several hundred seconds after core bounce, and have succeeded in explaining observational properties of individual SN events (e.g., SN 1987A; \citealt{2015ApJ...806..275P}), as well as the Galactic chemical evolution highlighted by the abundance patterns of extremely metal-poor (EMP) stars (e.g., \citealt{2019ApJ...870....2C}). 
Therefore, these simulations support that the neutrino-driven model is promising as a standard explosion mechanism of CCSNe.

However, it is unclear how the results in `artificial-explosion' simulations can overlap the nature of the explosion found in the ab-initio simulations, due to the complexity of the physics to be calibrated. 
Indeed, a rational strategy would be first to identify the major differences, if they would exist, between the `artificial-explosion' models and the `ab-initio' simulations, and then to investigate a key ingredient that might be still missing in the ab-initio simulations. 
In this respect, many lessons can be obtained by studying the nature of the classical `thermal bomb' or `piston' explosion models (e.g., \citealt{1995ApJS..101..181W}; \citealt{1996ApJ...460..408T}; \citealt{2008ApJ...673.1014U}), 
which form a basic background of the SN nucleosynthesis and have been extremely successful to explain many observational data for SNe and the chemical evolution.

It has been suggested that the successes of the classical thermal bomb/piston models and the calibrated neutrino-driven models would be mainly attributed to the rapid nature of the explosion in these simulations \citep{2019MNRAS.483.3607S}. 
The classical models assume that the energy injection is nearly instantaneous. 
The `artificial-explosion' simulations also show that the explosion energy grows `rapidly', especially at the initiation of the explosion; this is required as the outcome of the `calibration' (e.g., to explain the explosion energy of individual SNe). 
The timescale, in which the explosion energy reachs to $10^{51}$ erg, estimated from the linear extrapolation is about $\lesssim300$ msec in these models. 
On the other hand, most, if not all, of the ab-initio simulations represent an explosion in which the observed explosion energy ($\sim 10^{51}$ ergs) is reached in a timescale of at least $\sim$1 second, or even longer 
(see Table.1 of \citealt{2019MNRAS.483.3607S}; e.g., \citealt{2009ApJ...694..664M}; \citealt{2010PASJ...62L..49S}; \citealt{2012ApJ...749...98T}; \citealt{2012ApJ...756...84M}; \citealt{2016ApJ...818..123B}; \citealt{2015PASJ...67..107N}). 
Namely, the current ab-initio simulations predict the `slow explosion', while the artificially-tuned models to satisfy some observational constraints require the `rapidly' explosion.

However, little attention has been paid to the impact of the explosion timescales on the nucleosynthesis products; 
in this paper, we aim to show that a key ingredient leading to the success of the artificially calibrated models to explain various observations data is the explosion timescale. 
There are suggestions that the products via complete silicon burning, especially $^{56}$Ni, is sensitive to the energy deposition rate rather than the total explosion energy itself; 
there is a tendency that the slow explosion model suppresses the production of $^{56}$Ni (\citealt{2009MNRAS.394.1317M}; \citealt{2015MNRAS.451..282S}; \citealt{2019MNRAS.483.3607S}). 
There were however a few limitations in these studies. 
(1) In these previous studies, the mass of synthesized $^{56}$Ni was crudely estimated without performing detailed nuclear reaction network calculations. 
(2) the hydrodynamic behavior and various uncertainties related to the progenitor mass and the so-called mass cut (i.e., the boundary which separates the collapsing remnant and the SN ejecta) were not examined except for \cite{2019MNRAS.483.3607S}. 
Additionally, the effects and uncertainties of the finite growth timescale have been studied by \cite{2007ApJ...664.1033Y} and \cite{2015ApJ...814...63M}; 
however, \cite{2007ApJ...664.1033Y} represents an extreme black-hole forming SN and thus their results cannot be immediately compared to observations of typical CCSNe or the chemical evolution. 
The model of \cite{2015ApJ...814...63M} represents a steep energy growth, especially at the initiation of the explosion, and considers the amount of $^{56}$Ni as a parameter.
In this paper, by performing the nuclear reaction network calculation for a range of the explosion timescale, from the rapid to slow models, we aim at providing nucleosynthetic diagnostics on the explosion timescale. 
In doing this, we compare abundances of various elements/isotopes found in the models with those inferred from various observations.

In this paper, we simulate one-dimensional hydrodynamic and nucleosynthesis above the proto-neutron star core, by parameterizing the nature of the explosion mechanism by the energy growth timescale 
($t_{\rm grow}$; the timescale in which the explosion energy is reached to $10^{51}$ ergs since the initiation of the explosion). 
We compare our numerical results to various observational constraints; the masses of $^{56}$Ni derived for typical CCSNe, the masses of $^{57}$Ni and $^{44}$Ti observed for SN 1987A, and the abundance patterns observed in the EMP stars. 
By these comparisons, we discuss the appropriate energy growth timescale for typical CCSN explosion mechanism, i.e., an important constraint on the nature of the explosion. 
In Section \ref{ sec:model } we describe the treatment of two important ingredients in our modeling; the energy growth timescale and the mass cut. 
In Section \ref{ sec:method }, we describe our simulation framework, the progenitor model, and post-processing analysis. 
Our results are summarized in Section \ref{ sec:result }, followed by detailed comparisons to various observational data in Section \ref{ sec:observation }.
Conclusions are presented in Section \ref{ sec:discussion }.

\section{Models  }\label{ sec:model }

\subsection{Explosion Models }\label{ sec:grow }
In order to mimic the results of ab-initio multi-D simulations for a standard neutrino-driven delayed explosion, 
we consider a situation where the SN  explosion is energized by a continuous energy input at the central core, rather than an instantaneous energy input. 
Additionally, since the purpose of this study is to clarify the impact of the energy growth timescale $t_{\rm grow}$ on nucleosynthesis, our simulation needs to be able to manipulate a broad range of energy growth timescales explicitly. Therefore, our simulation is opted to remove the proto-neutron star (PNS) core and drive an explosion by injecting energy at the PNS surface.

\begin{figure}[ht]
\begin{center}
  \includegraphics[width=80mm]{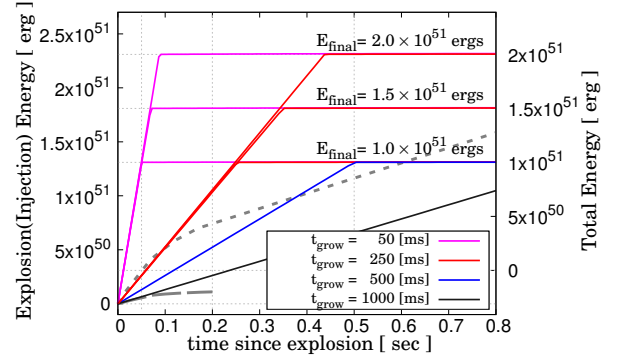}
  \caption{The relation between the definition of $t_{\rm grow}$ and the final energy $E_{\rm final}$. 
The models having same $t_{\rm grow}$ are shown by the same color; the magenta line is for $t_{\rm grow}=$ 50 ms, the red line for 100 ms, the blue line for 500 ms, and the black line for 1000 ms. 
The gray dotted and dashed lines are the time evolution of the explosion energy by \cite{2016ApJ...821...38S} (`artificial' explosion) and that by \cite{2016MNRAS.461L.112T} (ab-initio explosion), respectively.
Note that $E_{\rm bind}$ is determined from the pre-SN star and the figure is for $M_{\rm ZAMS}=$ 20.0 $M_\odot$ model.}
  \label{ fig:energy } 
\end{center}
\end{figure}

Here, the input energy is assumed to increase linearly, i.e., the energy input rate is taken to be constant. 
The energy input is terminated when a desired explosion energy ($E_{\rm exp}$) is reached. 
This prescription is motivated by the result of the recent ab-inito simulations, 
where the diagnostic energy is found to increase linearly to the first approximation (e.g. \citealt{2016MNRAS.461.3296N}). 
The constant energy input rate is thus modeled as follows:  
\begin{align}
\dot{E}_{\rm exp}&=E_{\rm exp}/t_{\rm exp}=(E_{\rm final}+|E_{\rm bind}|)/t_{\rm exp}\label{eq:def1}~,\\
t_{\rm exp}&=\frac{E_{\rm exp}}{E_{\rm ref}+|E_{\rm bind}|}\times t_{\rm grow} \label{eq:def2}~,
\end{align}
where $E_{\rm final}$ is the final energy of the supernova, and $E_{\rm exp}$ is the total injected energy. 
$t_{\rm grow}$ is defined as the energy growth timescale in which the explosion energy (i.e., the injected energy subtracted by the bound energy) is reached to the canonical energy of the explosion in normal CCSNe, which is defined as $E_{\rm ref}\equiv 10^{51}$ ergs. 
Therefore, equations (\ref{eq:def1}) and (\ref{eq:def2}) show that $t_{\rm exp}$ corresponds to the energy growth timescale up to $E_{\rm final}$ (Figure \ref{ fig:energy }).  
In this paper, we treat $t_{\rm grow}$ and $E_{\rm final}$ as free parameters which reflect the nature of the explosion mechanism. 
The detail method of how the energy is injected is described in \S \ref{ sec:method }.

\subsection{Treatment of the mass cut }\label{ sec:cut }
In the real CCSN explosion, a fraction of materials in the deepest layer may remain gravitationally bound and fallback onto the compact remnant. 
This fallback effect, however, cannot be directly computed in the 1D thermal-bomb or piston driven models. 
Nevertheless, the ejection of even small amounts of iron core material results in a broad variation in the abundance patterns (e.g., \citealt{1996snih.book.....A}).
In order to capture this fallback effect, traditional SN nucleosynthesis studies adopt a prescription called the `mass cut', which artificially separates the ejecta from the collapsing core. 
Since the 1D nucleosynthesis models commonly show a strong variation of the neutron excess as a function of depth near the Fe-core surface, 
the resulting abundance patterns, especially those of neutron-rich Fe-peaks, are sensitive to the choice of the mass cut (e.g., \citealt{1996ApJ...460..408T}). 
Inversely, the mass cut can be chosen phenomenologically as guided by a particular set of observational constraints.  

Here we use the following criteria to constrain the mass cut position: 
(1) The `deep ejecta model'; all the materials above the outermost edge of the energy injection region are ejected. 
The deeper mass cut leads to a larger amount of newly synthesized species ejected at the explosion; 
therefore this model is regarded as the maximum limit in terms of the amount of newly synthesized elements ejected by the explosion.
Since our constraints in the following sections are mostly on the capability of the models to produce a sufficient amount of nucleosynthesis products, 
this choice allows a conservative constraint on $t_{\rm grow}$.

To test the uncertainty associated with the choice of the mass cut, we also examine the following case: 
(2) The `EMP ratio model'; the masscut is set to reproduce the Ni/Fe ratios observed in the EMP stars.
As described in \S \ref{ sec:MPS }, it can be assumed that the EMP stars preserve individual CCSN abundance patterns.
Therefore, the abundance patterns of the EMP stars have been adopted to constrain the nature of population III or II massive stars and their SN explosions (e.g., \citealt{2007ApJ...660..516T}; \citealt{2010ApJ...724..341H}). 
We follow this approach and require that the CCSN ejecta must match the abundance patterns of the EMP stars.
However, we note that it is difficult to fine-tune all the abundance patterns of the EMP stars at a single mass cut position \citep{2003Natur.422..871U}. 
In this study, therefore, we consider the range of the abundance patterns obtained by the two treatments of the masscut reflects the uncertainty.
The detail on the result related to the choice of the mass cut is described in \S \ref{ sec:mass-cut }.

\section{numerical method}\label{ sec:method }
\subsection{Numerical set-up}\label{ sec:setup }
To simulate the explosion, we employ a 1D Lagrangian code based on 
{\tt blcode}\footnote{https://stellarcollapse.org/SNEC}, 
which solves Newtonian hydrodynamics and is a prototype code of {\tt SNEC}\citep{2015ApJ...814...63M}.
Basic equations under spherically symmetric configuration as we perform in this paper are given as follows: 
\begin{align}
\cfrac{\partial r}{\partial M_{r}}&=\cfrac{1}{4\pi r^2\rho} \label{eq:basic1}~,\\
\cfrac{D v}{D t}&=-\cfrac{GM_{r}}{r^2}-4\pi r^2\cfrac{\partial P}{\partial M_{r}} \label{eq:basic2}~,\\
\cfrac{D \epsilon}{D t}&=-P\cfrac{D}{D t}\left(\cfrac{1}{\rho}\right)+\dot{S} \label{eq:basic3}~,
\end{align}
where $r$ is radius, $M_{r}$ is mass coordinate, $t$ is time, $\rho$ is density, $v$ is radial velocity, $P$ is pressure, $\epsilon$ is specific internal energy, $\dot{S}$ is the local energy generation rate per unit mass, and $D/Dt \equiv \partial/\partial t + v_r\partial/\partial r$. Artificial viscosity by \cite{1950JAP....21..232V} is employed to capture a shock. 
The system of equations (\ref{eq:basic1})-(\ref{eq:basic3}) is closed with the Helmholtz equation of state \citep{2000ApJS..126..501T}, which describes the stellar plasma as a mixture of arbitrarily degenerate and relativistic electrons and positrons, blackbody radiation, and ideal Boltzmann gases of a defined set of fully ionized nuclei, taking into account corrections for the Coulomb effects. 
It includes a nuclear burning to follow the energy generation, by solving a 21 $\alpha$-isotope reaction network\footnote{http://cococubed.asu.edu/code\_pages/codes.shtml} which is derived from \cite{1978ApJ...225.1021W}; 
neutron, proton, $^1$H, $^2$H, $^3$He,
$^4$He, $^{12}$C, $^{14}$N, $^{16}$O, $^{20}$Ne, $^{24}$Mg ,$^{28}$Si, $^{32}$S, $^{36}$Ar, 
$^{40}$Ca, $^{44}$Ti, $^{48}$Cr, $^{52}$Fe, $^{54}$Fe, $^{56}$Cr, $^{56}$Fe, $^{56}$Ni. 
For a more accurate assessment of the nucleosythesis, a post-processing analysis is performed with a nuclear reaction network including 640-nuclear species as described in \cite{1999ApJS..124..241T}.

\begin{figure}[ht]
\begin{center}
  \includegraphics[width=80mm]{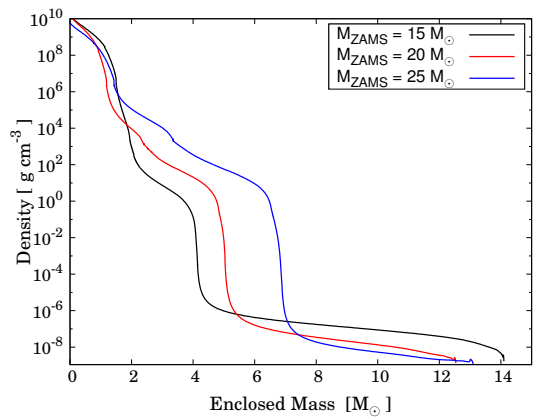}
  \caption{Density profiles vs. enclosed mass for the progenitor models with $M_{\rm ZAMS}=$ 15.0 $M_\odot$ (black line), 20.0 $M_\odot$ (red line), and 25.0 $M_\odot$ (blue line).}
  \label{ fig:progenitor } 
\end{center}
\end{figure}

The pre-explosion structure as an input to our simulations is solar-metallicity, non-rotating stellar models obtained by the stellar evolution code {\tt MESA} \citep{2015ApJS..220...15P}. 
Since the nucleosynthesis from CCSNe can vary significantly across the ZAMS mass range, we generate the 3 pre-explosion models by evolving a zero-age main sequence (ZAMS) star, with the initial mass of 15.0, 20.0, and 25.0 $M_\odot$, to the point of the iron core collapse.  
The final stellar mass is $\approx$14 $M_\odot$ in all models.
These models cover a range of the initial masses for different type of SNe, including the progenitor of SN 1987A (e.g., \citealt{2005AstL...31..806U}).
We note, however, that our simulation is not tuned to model any peculiar SNe like SN 1987A. 
Figure \ref{ fig:progenitor } shows the density profiles of the progenitor models.

In all the models, the computational domain extends from a minimum radius near the outer edge of the PNS (at the enclosed mass of $M_{r, {\rm min}}\approx1.1-1.3 M_\odot$) up to the photosphere of  the progenitors (at $M_{r, {\rm max}}\approx14 M_\odot$). 
The grid has logarithmic spacing, with 2000 cells in the radial direction. 
The minimal radius is determined by the location where the electron mass fraction is ${\rm Ye}=0.48$., which we assume to represent the edge of the PNS\footnote{The inner boundary here is set as deep as possible to provide conservational constraints. This does not necessarily correspond to the actual PNS surface (\S \ref{ sec:mass-cut }).}. 
In order to check the impact of this choice, we additionally performed simulations where the inner boundary is deeper or shallower by 0.05 $M_\odot$, and we found no significant difference in the hydrodynamic behaviors from our reference models. 
At the inner boundary, we opted to remove the PNS and drive an explosion by injecting energy in the innermost zones (\S\ref{ sec:boundary } for details).

\subsection{Method of injecting the energy}\label{ sec:boundary }
Our simulations are performed by injecting the energy into the innermost 20 zones ($\equiv  0.005 M_\odot$, at the outer edge of the PNS). 
The energy growth timescale $(t_{\rm grow})$ and the final energy $(E_{\rm final})$, which correspond to the duration and magnitude of these artificial explosions, are treated as free parameters.
The energy growth timescale is converted to $t_{\rm exp}$ by eq.(\ref{eq:def2}) and the energy is injected at a constant late $\dot{E}_{\rm exp}=E_{\rm exp}/t_{\rm exp}$ for $t \leqq t_{\rm exp}$. 
To further reduce the number of free parameters, we assume that the energy injection at the innermost region is dominated by the thermal energy (so-called `thermal bomb'). 
We set $\dot{E}_{\rm kin}/\dot{E}_{\rm exp}\leqq 0.01$ (where $\dot{E}_{\rm kin}$ indicates the kinetic energy content). 
We note that the energy equitation is quickly reached behind the shock, and thus the result would not be sensitive to the ratio of kinetic energy content; the thermal bomb and piston models provide similar results \citep{2007ApJ...664.1033Y}.

\begin{figure}[ht]
  \begin{center}
	  \includegraphics[width=80mm]{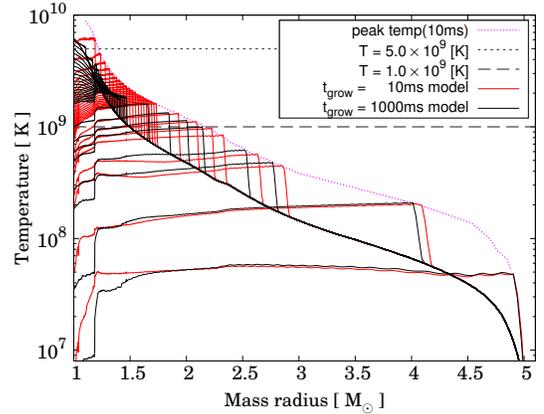}
  \caption{Temperature profiles at various times, for two different values of $t_{\rm grow}$.
Each snapshot corresponds to roughly a doubling time, e.g., $\sim$10 ms, $\sim$20 ms, $\sim$40 ms, up to $\sim$130 seconds. 
Red line: explosion with $t_{\rm grow}$=10 [ms], $M_{\rm ZAMS}=20M_\odot$, and $E_{\rm final}=1.0\times10^{51}$ erg.
Black line: the same model except for $t_{\rm grow}$=1000 [ms].
The shock wave is successful to propagate through the stellar core to the outer layer for both $t_{\rm grow}$. 
Also, it was confirmed that the behavior of the shock wave in the outer layer does not depend on $t_{\rm grow}$.}
  \label{ fig:temperature } 
  \end{center}
\end{figure}

To limit a range of a possible value for $t_{\rm grow}$, we impose the condition that the pressure of the injected materials ($P_{\rm exp}$, which is a function of $t_{\rm grow}$) must overcome the ram pressure of the infalling progenitor material ($P_{\rm ram}$), i.e.,
\begin{equation}\label{eq:ram}
P_{\rm exp} >P_{\rm ram}=\rho_{\rm pg} v^2_{\rm pg} \ ~,
\end{equation}
Here, $\rho_{\rm pg}$ and $v_{\rm pg}$ are the density and velocity of the pre-collapse initial values for the progenitor material at the inner boundary where the energy is injected.
For the set of our model parameters, 
equation (\ref{eq:ram}) leads to the following criterion on $t_{\rm grow}$; $t_{\rm grow}<2500$ ms. 
Therefore, we select $t_{\rm grow}=$10, 50, 100, 200, 250, 400, 500, 1000, 2000 [ms] in our simulations. 
Finally, in the selected range of $t_{\rm grow}$, we confirmed that the shock wave propagates through the stellar core to the outer layer. 
Fig.\ref{ fig:temperature } shows the time evolution of the shock wave with $t_{\rm grow}=$ 10 and 1000 ms. 
It is confirmed that the behavior of the shock wave in the outer layer does not depend on $t_{\rm grow}$.

To check the effect of the diversity in $E_{\rm final}$, we investigate $E_{\rm final}=1.0-2.0\times E_{\rm ref}(\equiv1.0-2.0\times 10^{51}$ ergs) for the $20.0M_\odot$ pre-explosion model.
The nature of the explosion, particularly in the nucleosynthesis, may be sensitive to $E_{\rm final}$, and the range of $E_{\rm final}$ can be broad in observed canonical CCSNe, from 0.8 to 3.0 $\times 10^{51}$ergs (\citealt{2003IAUS..212..395N}). 
Note that $t_{\rm grow}$ is defined in a way so that it is independent from $E_{\rm final}$, since this is the timescale in which the explosion energy is reached to $E_{\rm ref}\equiv 10^{51}$ ergs. 

\section{results}\label{ sec:result }
\subsection{Dynamics and Abundance distribution}\label{ sec:result2 }
\subsubsection{Temperature profiles}\label{ sec:temp }
\begin{figure}
  \begin{center}
	  \includegraphics[width=80mm]{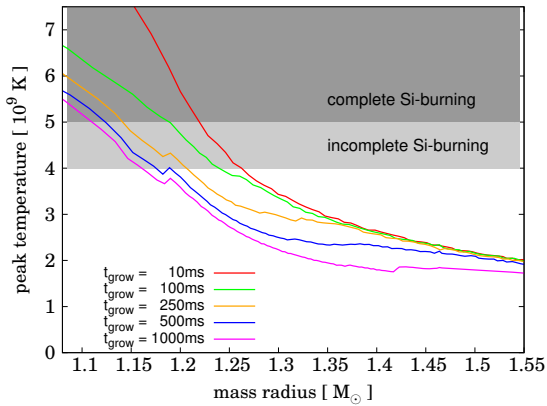}
  \caption{The peak temperature evolution behind the shock wave as a function of the enclosed mass $M_r$, for the energy growth timescale $t_{\rm grow}$=10, 100, 250, 500, 1000 msec, for the model with $M_{\rm ZAMS}=20M_\odot$ and $E_{\rm final}=1.0\times10^{51}$ ergs. The dark-grey region corresponds to $T>5\times10^9$ K (complete Si-burning region, which is one of the criteria to synthesize $^{56}$Ni), and the light-grey region to $4\times10^9<T<5\times10^9$ K (incomplete Si-burning region).}
  \label{ fig:temp } 
  \end{center}
\end{figure}
In order to describe how hydrodynamical behaviors are affected by the energy growth timescale, we will use the model with $M_{\rm ZAMS}=20M_\odot$ and $E_{\rm final}=1.0\times10^{51}$ ergs as an example throughout \S \ref{ sec:result }. 
How $M_{\rm ZAMS}$ and $E_{\rm final}$ affect the outcomes will be discussed in \S \ref{ sec:observation }.
Figure \ref{ fig:temp } shows the temperature evolution just behind the shock wave, for $t_{\rm grow}$=10, 100, 250, 500, 1000 ms. 
The model with $t_{\rm grow}$=10 ms is set up to mock up an instantaneous explosion, while the models with $t_{\rm grow}\geqq$500 ms represent a slow explosion. 
There is a clear difference in the mass coordinate that the shock wave sweeps until the temperature decreases below $T \sim 5\times10^9$ K (shown by the dark gray region). 
This region undergoes the complete Si-burning (hereafter the complete Si-burning region).
In the instantaneous explosion model ($t_{\rm grow}$=10 ms), a strong shock propagates up to $M_r \approx 1.22M_\odot$ keeping a high temperature ($T>5\times10^9$ K) to trigger the complete Si-burning. 
On the other hand, in the slow explosion model (e.g., $t_{\rm grow}$=1000 ms), the shock is weak, and the temperature decreases to $T<5\times10^9$ K before a shock reaches to $M_r\approx1.10M_\odot$. 
That is, it is estimated that the complete Si-burning product in the slow explosion model would be by $\sim0.1M_\odot$ less than that in the instantaneous explosion model.

This can be understood by the following simplified analytic estimate. 
For simplicity, we consider that a uniform fireball is created behind the shock wave. 
We further assume spherical symmetry and a constant expansion (shock) velocity ($v_{\rm shock}$). 
The post-shock region is dominated by radiation and contains the energy $\dot{E}_{\rm exp}\cdot t\leqq E_{\rm exp}$. 
Then, the following approximate relation holds for the properties of the shocked region:
\begin{equation}
  \cfrac{4\pi}{3}~aT^4r_{\rm sh}^3(t) = \begin{cases}
    \dot{E}_{\rm exp}\cdot t & (t \leqq  t_{\rm grow}) ~, \\
    E_{\rm exp} & (t >  t_{\rm grow}) ~,
  \end{cases}
\end{equation} 
where $r_{\rm sh}(t)=R_{\rm Ye=0.48}+v_{\rm shock}\cdot t$ is the shock radius, and $R_{\rm Ye=0.48}$ is the radius of the inner boundary (set by Ye$=0.48$). 
Adopting $R_{\rm Ye=0.48}=10^8$ cm and $v_{\rm shock}={\rm const.}=10^9$ cm s$^{-1}$, 
we estimate that the shock wave can expand to $r_{\rm sh} \sim 3.7\times 10^8$ cm with $T>5\times10^9$ K, for the case of $t_{\rm grow}=10$ ms, and this estimate is consistent with \cite{2002RvMP...74.1015W}, in which the similar expression is derived for an instantaneous energy deposition. 
This estimate suggests that the complete Si burning is active until the shock radius reaches to $\sim3.7\times10^8$ cm, that is, $\sim270$ ms after the initiation of the explosion.
Moreover, if $t_{\rm grow}\geqq$500 ms, this radius is reduced to $ r_{\rm sh}<2.4\times 10^8$ cm, which is smaller than the case of the instantaneous explosion.
From this simple analytic relation, it is seen that $t_{\rm grow}\sim $270 ms is the criterion above which the result becomes different from that found in the instantaneous explosion model, which has been successful to explain many observational data for SNe and the chemical evolution (e.g., \citealt{1995ApJS..101..181W}).

\subsubsection{Abundance distribution}\label{ sec:abundance } 
\begin{figure*}
\gridline{\fig{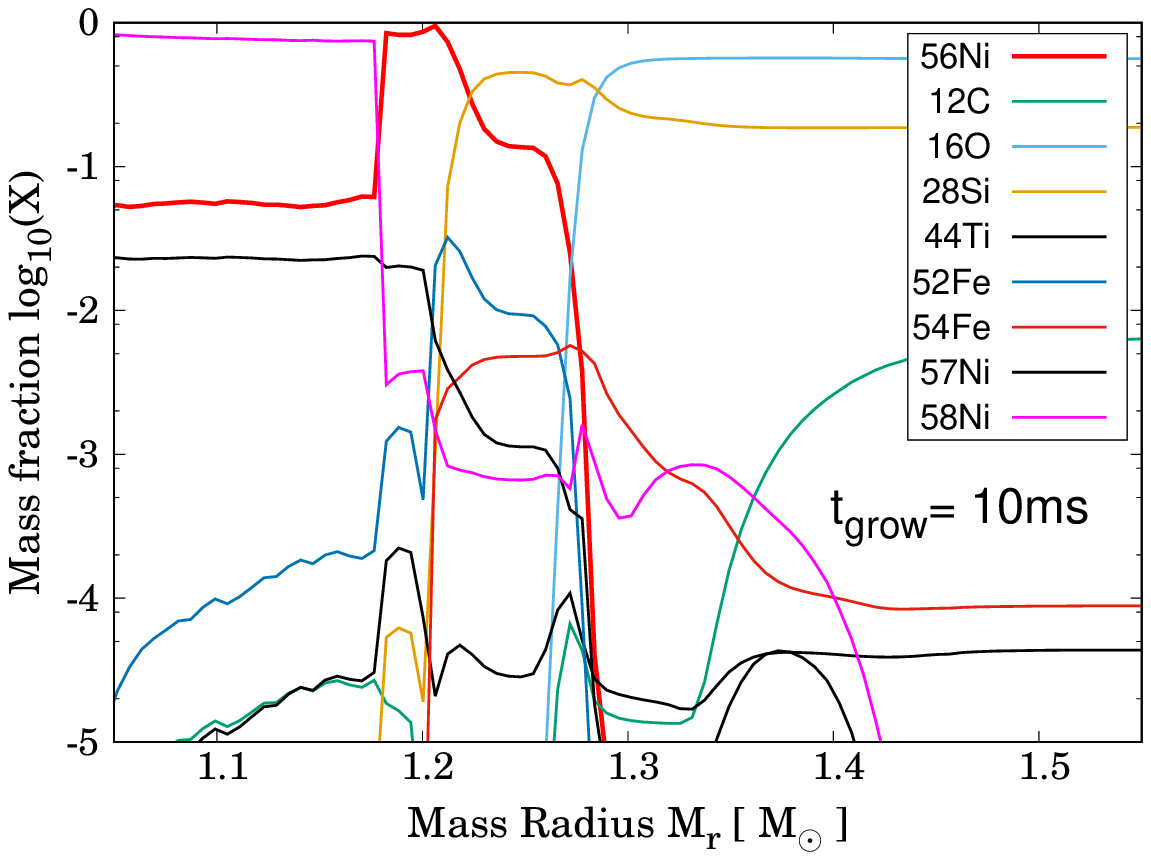}{0.33\textwidth}{(a)}
          \fig{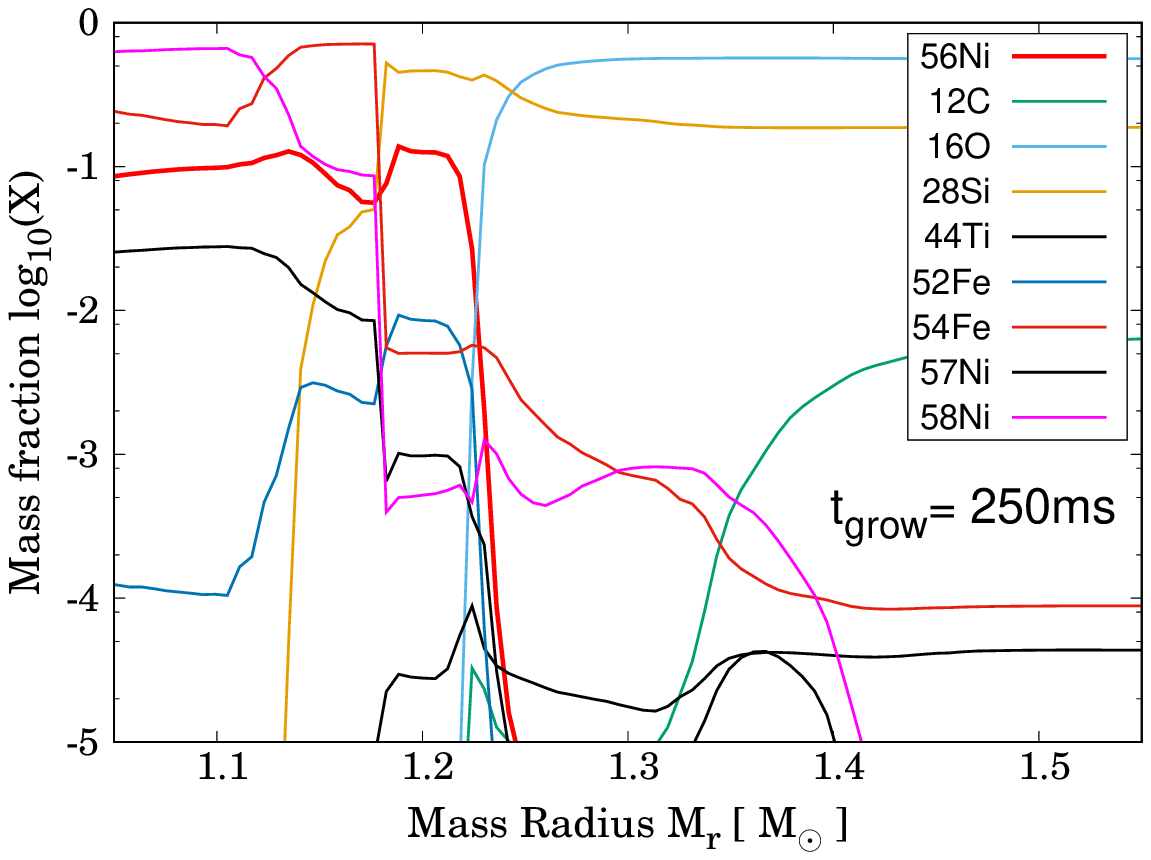}{0.33\textwidth}{(b)}
          \fig{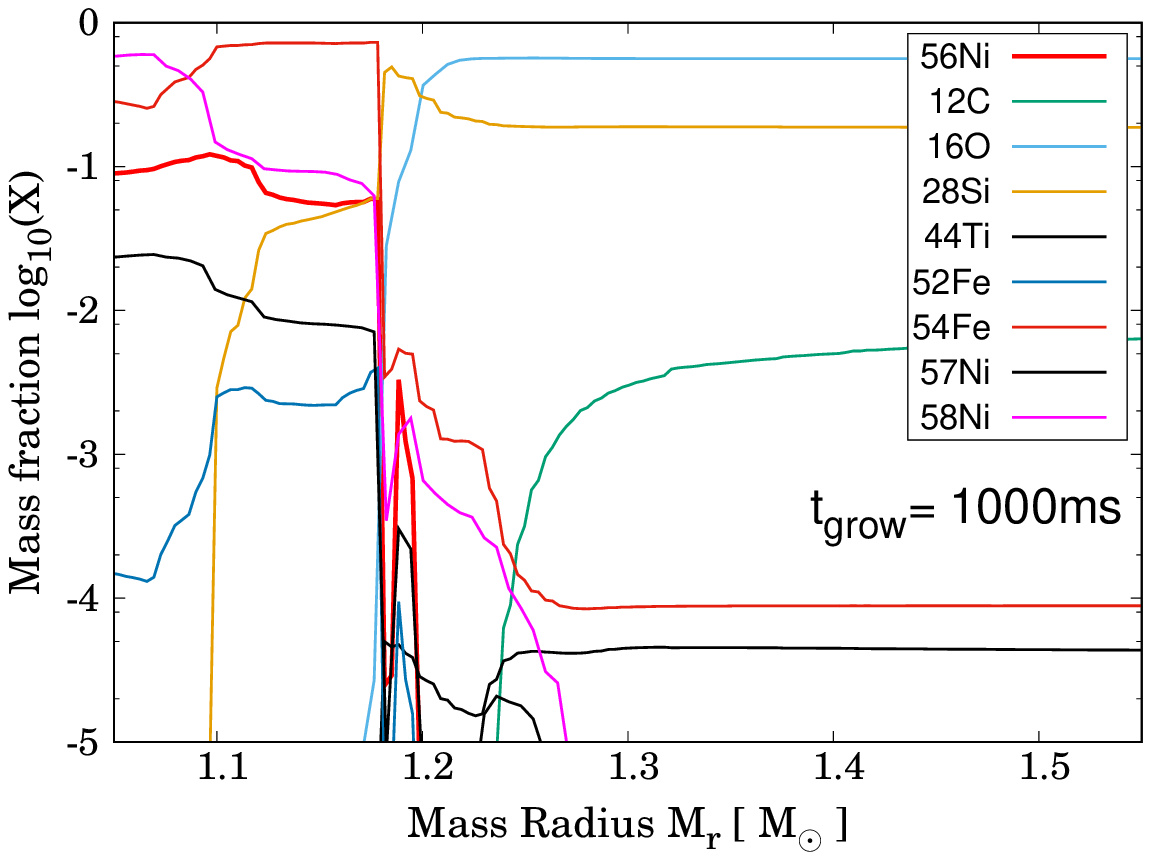}{0.33\textwidth}{(c)}
          }
  \caption{Abundance distribution as a function of the enclosed mass $M_r$, for $t_{\rm grow}$= (a) 10 msec, (b) 250 msec, and (c) 1000 msec.
All the models here are with 20$M_\odot$ and $E_{\rm final}=1.0\times10^{51}$ ergs.}
  \label{ fig:abundance } 
\end{figure*}

Fig.\ref{ fig:abundance } shows the abundance distribution as a function of the enclosed mass $M_r$ for $t_{\rm grow}$=10, 250 and 1000 msec, with $M_{\rm ZAMS}=20M_\odot$ and $E_{\rm final}=1.0\times10^{51}$ ergs. 
The characteristic burning region can be reviewed by the distribution of $^{56}$Ni, $^{57}$Ni, and $^{28}$Si. 
The model with $t_{\rm grow}$=10 ms shows the following; 
$(a)$ $^{56}$Ni and $^{57}$Ni are predominantly synthesized up to $M_r \sim1.22M_\odot$, 
where the peak temperature reaches to $T>5\times10^9$ K (see Fig.\ref{ fig:temp }).
This region, which contains $\sim0.2 M_\odot$, undergoes the complete Si-burning. 
$(b)$ $^{28}$Si is abundantly synthesized at the incomplete Si-burning region, which extends from $1.22M_\odot$ to $\sim1.3M_\odot$ outside the complete Si-burning region.
$(c)$ Beyond $1.3M_\odot$, the pre-SN abundance is largely conserved (no-burning region).

The distributions of these different layers are affected by $t_{\rm grow}$. 
The model with $t_{\rm grow}$=250 ms shows the following; 
$(a)$ The complete Si burning region is narrower than the model with $t_{\rm grow}$=10 ms, and extends only up to $M_r\sim1.15M_\odot$.
$(b)$ The incomplete Si-burning region extends to $M_r\sim1.22M_\odot$. 
These distributions are consistent with the regions which reach to $T>5\times 10^9$ K and $T>4\times 10^9$ K, respectively, in the model with $t_{\rm grow}$ = 250 ms as seen in Figure \ref{ fig:temp }.
$(c)$ Beyond $1.25M_\odot$, the pre-SN abundance is largely conserved.

The abundance distribution is further changed dramatically for the slowest model with $t_{\rm grow}$=1000 ms; 
$(a)$ there is virtually no complete Si-burning region ($M_r\lesssim1.1M_\odot$), and $(b)$ even the incomplete Si-burning region shrinks in the mass coordinate drastically, covering only up to $M_r\sim1.17M_\odot$. 
Namely, the explosive nucleosynthesis is strongly suppressed due to a decrease in the peak temperature (see Fig.\ref { fig:temp } and section \ref{ sec:temp }) for the slowest explosion model. 
Consequently, $(c)$ almost all the ejecta ($M_r\gtrsim 1.17M_\odot$) are composed of pre-SN abundance for the slowest model ($t_{\rm grow}$=1000 ms). 
In \S \ref{ sec:mass-cut }, the dependence of the amount of iron peak element on the mass cut position will be discussed based on the results of nucleosynthesis.

\subsection{Mass cut position}\label{ sec:mass-cut }
\begin{figure}
\begin{center}
  \includegraphics[width=80mm]{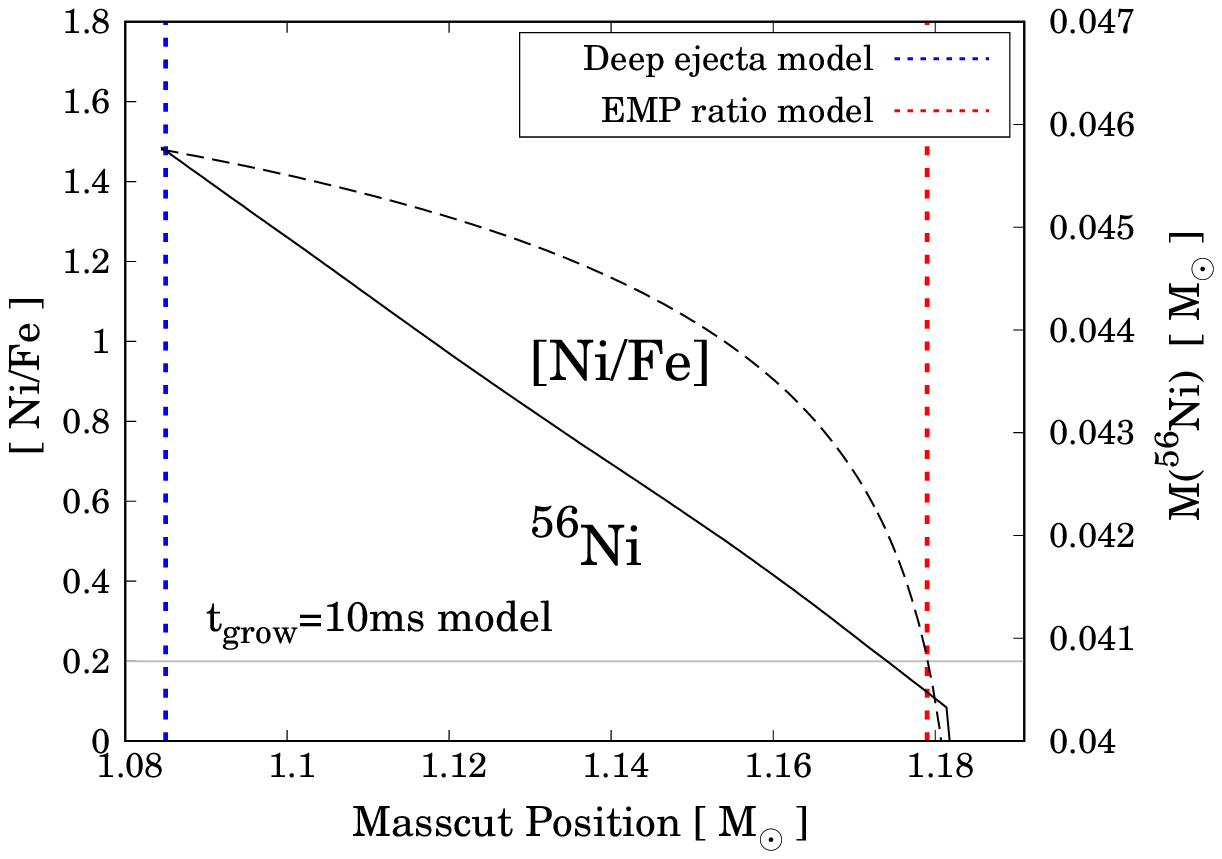}
  \includegraphics[width=80mm]{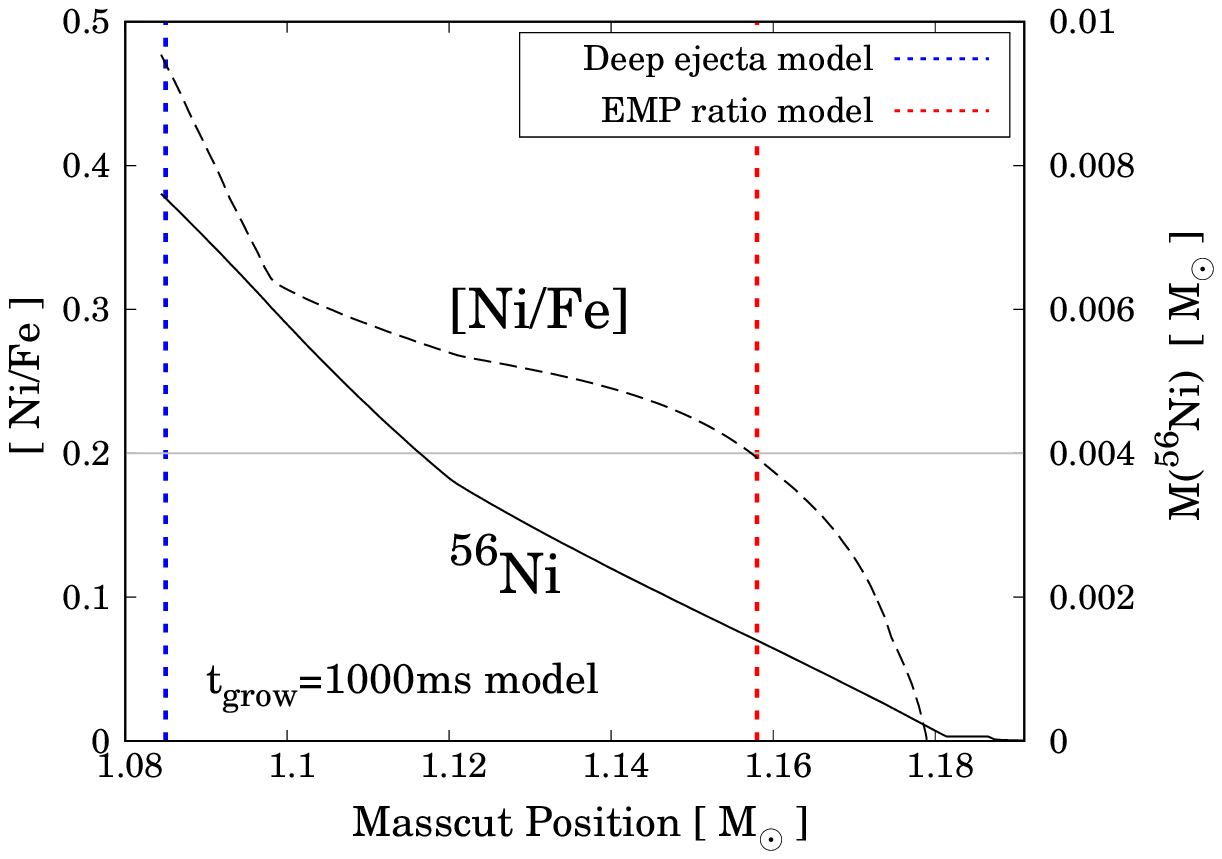}
  \caption{Dependence of the abundance ratio [Ni/Fe] and the ejected amount of $^{56}$Ni on the mass coordinate at the mass cut $M_{\rm cut}$, for $t_{\rm grow}$=10 and 1000 msec with $M_{\rm ZAMS}=$20$M_\odot$ and $E_{\rm final}=1.0\times10^{51}$ ergs. 
The gray solid lline shows [Ni/Fe]=0.20, corresponding to the abundance ratio observed in the EMP stars (e.g., \citealt{2004A&A...416.1117C}). 
The blue and red dotted lines corresponds the mass cut positions in the `Deep ejecta model' and the `EPM ratio model', respectively (see section.\ref{ sec:cut }). 
For example, in the `Deep ejecta model' with $t_{\rm grow}$= 10 msec, the ejecta have [Ni/Fe]$\approx$1.5 and $M ({\rm {}^{56}Ni})\approx0.0455M_\odot$ .}
  \label{ fig:mass-cut } 
\end{center}
\end{figure}
In this study, the fallback of matter onto the compact remnant, which affects the mass and composition of the ejecta, is artificially modeled as a `mass cut', as described in section \ref{ sec:cut }.
The uncertainty related to the mass cut is taken into account in the subsequent section \ref{ sec:observation }, where we compare the amount and ratio of each isotope in the ejecta with observations.
Here, we discuss the dependence of yields on the mass cut position, $M_{\rm cut}$. 
Then, we describe the result of the mass cut position which is constrained phenomenologically with the nucleosynthesis through the Ni/Fe ratio within the ejecta. 
Note that our model assumes the solar metallicity. 
Therefore, in order to compare with the EMP stars, here we focus on the region that experienced explosive nuclear burning. 
By focusing only on this region and omitting the metal content which has already been contained at ZAMS (i.e., solar abundance).
We define the region that experienced explosive nuclear combustion by the condition that the peak temperature is $T>1.0\times10^9$ K.
According to this definition, the heavy-element abundance ratios [X/Fe] is given as follows;
\begin{equation}
  [{\rm X}/{\rm Fe}]\equiv \log \left(\cfrac{M({\rm X})_{T_{\rm peak}>1.0\times10^9{\rm K}}}{M({\rm Fe})_{T_{\rm peak}>1.0\times10^9{\rm K}}}\right) - \log \left(\cfrac{\rm X}{\rm Fe}\right)_\odot~,
\end{equation} 
where $M({\rm X})_{T_{\rm peak}>1.0\times10^9{\rm K}}$ is the mass of element under consideration in the region that satisfies $T_{\rm reak}>1.0\times10^9$K, and $({\rm X/Fe})_\odot$ is the corresponding ratio at the solar abundance.
As seen in Firure \ref{ fig:temp }, the region that satisfies $T>1.0\times10^9$K converges to $\approx2M_\odot$ regardless of $t_{\rm grow}$.

Figure \ref{ fig:mass-cut } shows the result of [Ni/Fe]
and the mass of $^{56}$Ni as a function of $M_{\rm cut}$ for $t_{\rm grow}$=10 and 1000 msec. 
This result assumes that all the materials outside the mass cut position are ejected.
The red and blue dotted lines in the figure correspond to the mass cut positions selected in the `EPM ratio model' and the `Deep ejecta model', respectively (\S \ref{ sec:cut }).
It can be seen that [Ni/Fe] increases as the mass cut is deeper (i.e., $M_{\rm cut}$ is smaller) for any $t_{\rm grow}$ in the figure. 
In the SN ejecta, (stable) Ni is dominated by $^{58}$Ni, and Fe is by $^{54}$Fe and the decay product of $^{56}$Ni. 
Therefore, the ratio of Ni/Fe is expressed roughly by the ratio of $^{58}$Ni /($^{56}$Ni +$^{54}$Fe) \citep{2015ApJ...807..110J}. 
As seen in Figure \ref{ fig:abundance }, $^{58}$Ni is synsthesised in the innermost region. 
On the other hand, $^{56}$Ni and $^{54}$Fe are synsthesised in the outer region.
Consequently, the fraction of [Ni/Fe] is larger if the mass cut is taken at a deeper position (i.e., $M_{\rm cut}$ is smaller).
Figure \ref{ fig:mass-cut } also shows that the amount of ejected $^{56}$Ni is larger if the mass cut is deeper.

In the case of the `EPM ratio model', the mass cut position is selected so that the ejecta reproduces the composition of Ni/Fe observed in the EPM stars ([Ni/Fe] = 0.2; e.g., \citealt{2004A&A...416.1117C}), as this is the constraint we impose.
Note that the distribution of the explosive nucleosynthesis products differs for a different value of $t_{\rm grow}$. 
Therefore, we adjusted the mass cut position to [Ni/Fe] = 0.2 for each energy growth timescale and for each progenitor models, as shown by the red dotted line in Figure \ref{ fig:mass-cut }. 
On the contrary, in the case of the `Deep ejected model'  (in which the mass cut position is chosen to be at the innermost boundary since it is assumed that all the materials outside the surface of PNS are ejected), [Ni/Fe] becomes about unity or even larger, as shown by the blue dotted line in Figure \ref{ fig:mass-cut }. 
In this study, we define the edge of the PNS as the location where the electron mass fraction is ${\rm Ye}=0.48$, which is the deepest position that can be taken as a masscut position.
The resulting value, [Ni/Fe] $\sim 1$ is too large to represent `normal' SN explosions, as this is inconsistent with the Galactic chemical evolution. 
Therefore, this deep ejected mass cut should be regarded to be an extreme case, which produces the maximally allowed amount of newly synthesized isotopes. 
In other words, this case provides an extremely conservative upper limit, when using the amount of nucleosynthesis products to constrain $t_{\rm grow}$. 
As described above, we should note that this masscut position in the `Deep ejected model' is taken as deep as possible and does not necessarily correspond to the edge of the actual PNS.
We summarize in Table \ref{ table:masscut } the mass cut position adopted for each combination of $M_{\rm ZAMS}$, $E_{\rm final}$, and $t_{\rm grow}$.

In the following discussion, we adopt the `Deep ejected model' in \S \ref{ sec:56Ni } \& \ref{ sec:87A } where we are interested in the maximum amounts of the elements/isotopes which can be ejected by each explosion. In \S \ref{ sec:MPS }, we use both options on the mass cut to evaluate the uncertainty, in comparing the model predictions to the abundance patterns of the EMP stars.

\begin{sidewaystable}[htb]
  \begin{center}
    \caption{Summary of the mass cut position\label{ table:masscut }}
    \begin{tabular}{cc|cccccccccc}\hline\hline
     $M_{\rm ZAMS}$ & $E_{\rm final}$ & ${M_{\rm Ye=0.48}}^{a}$ & ${M_{\rm cut,10}}^{b}$ 
& ${M_{\rm cut,50}}^{b}$ & ${M_{\rm cut,100}}^{b}$ & ${M_{\rm cut,200}}^{b}$ 
& ${M_{\rm cut,250}}^{b}$ & ${M_{\rm cut,400}}^{b}$ & ${M_{\rm cut,500}}^{b}$ 
& ${M_{\rm cut,1000}}^{b}$ & ${M_{\rm cut,2000}}^{b}$ \\
     $(M_\odot)$ & ($10^{51}$ erg) & $(M_\odot)$ & $(M_\odot)$ & $(M_\odot)$ & $(M_\odot)$ & $(M_\odot)$ & $(M_\odot)$ & $(M_\odot)$ & $(M_\odot)$ & $(M_\odot)$ & $(M_\odot)$ \\ \hline
     15 & 1.0 & 1.34 & 1.493 & 1.493 & 1.4877& 1.437 & 1.430 & 1.414 & 1.410 & 1.399 & 1.394 \\
     20 & 1.0 & 1.09 & 1.179 & 1.179 & 1.167 & 1.147 & 1.147 & 1.158 & 1.158 & 1.137 & 1.168 \\
     20 & 1.5 & 1.09 & 1.178 & 1.178 & 1.166 & 1.240 & 1.219 & 1.158 & 1.158 & 1.136 & 1.167 \\
     20 & 2.0 & 1.09 & 1.177 & 1.177 & 1.167 & 1.278 & 1.233 & 1.158 & 1.158 & 1.137 & 1.168 \\
     25 & 1.0 & 1.12 & 1.269 & 1.269 & 1.263 & 1.274 & 1.277 & 1.273 & 1.268 & 1.259 & 1.254 \\
\hline
    \end{tabular}
	\begin{tablenotes}
$^{a}$ $M_{\rm Ye=0.48}$ is adopted as the mass cut position in the `Deep ejected model'.\\ 
$^{b}$ $M_{\rm cut,X}$ corresponds to the mass cut position in mass coordinate ($M_r$), in the `EPM ratio model' for $t_{\rm grow}{\rm =X}$ ms.
\end{tablenotes}
  \end{center}
\end{sidewaystable}

\section{Comparison to Observations}\label{ sec:observation }
\subsection{{\rm ${}^{56}$Ni} produced in typical CCSNe}\label{ sec:56Ni }
\begin{figure}
\begin{center}
  \includegraphics[width=80mm]{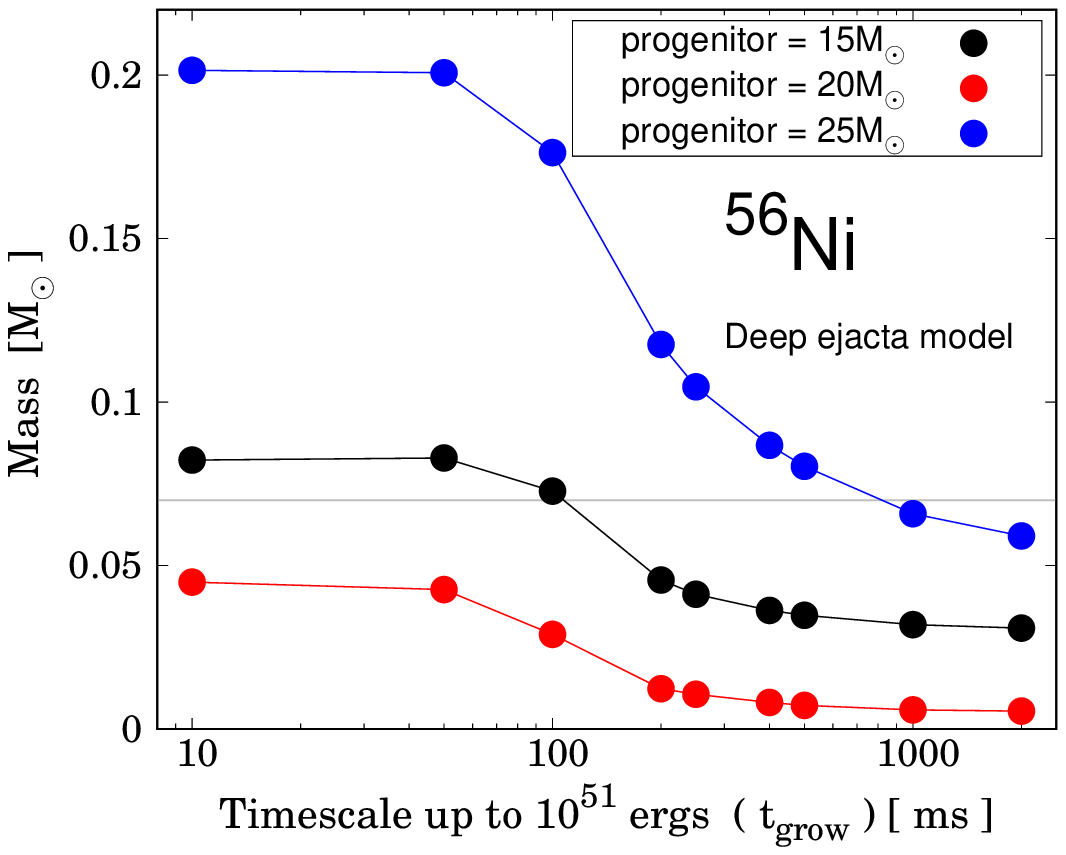}
  \includegraphics[width=80mm]{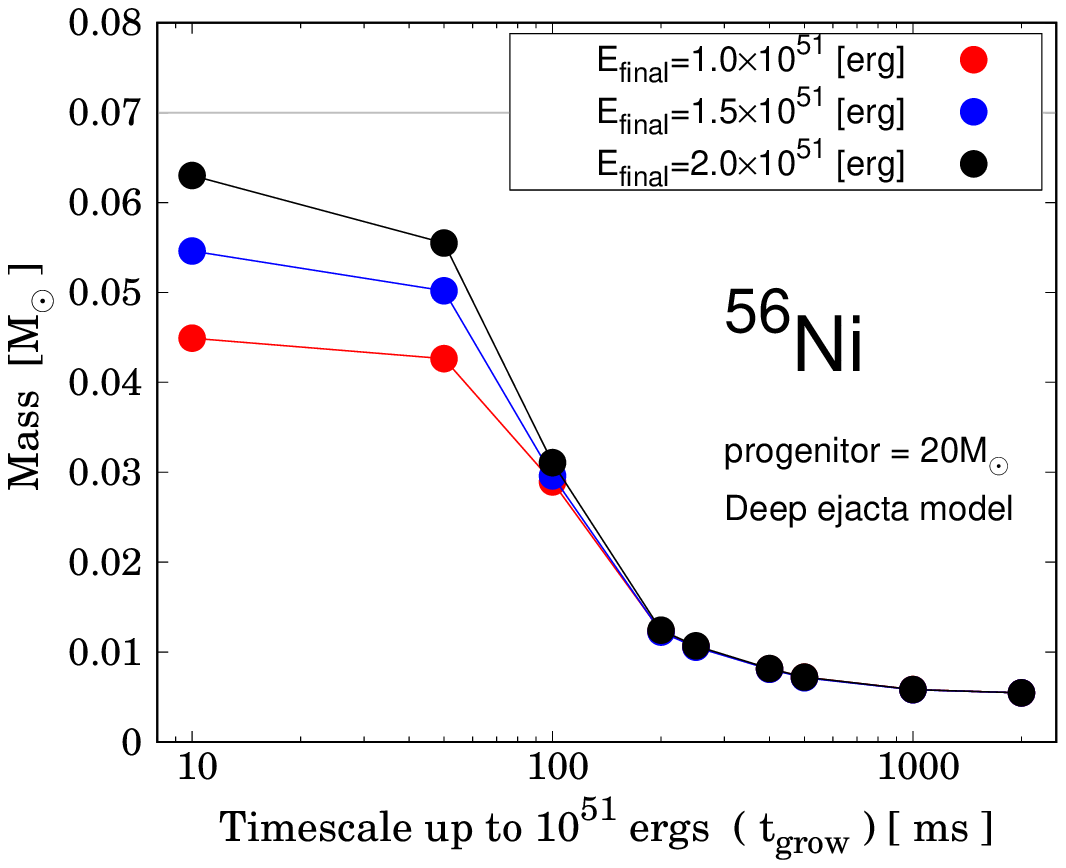}
  \caption{The energy growth timescale $t_{\rm grow}$ and the produced $^{56}$Ni mass. 
The left panel shows dependence on $M_{\rm ZAMS}$ (for given $E_{\rm final}=1.0\times10^{51}$ ergs), while the right panel shows dependence on $E_{\rm final}$ (for given $M_{\rm ZAMS}=20M_\odot$). The gray line corresponds to the typical amount of $M(^{56}{\rm Ni})=0.07M_\odot$ produced by CCSNe (e.g., SN 1987A, SN 1994I, SN 2002ap; \citealt{1989ARA&A..27..629A}; \citealt{1994ApJ...437L.115I}; \citealt{2002ApJ...572L..61M}). }
  \label{ fig:56Ni } 
\end{center}
\end{figure}

The amount of $^{56}$Ni ejected in individual SNe, which drives supernova brightness, is an important diagnosing indicator of the supernova explosion. 
It can be measured for many supernovae through light curve analyses with reasonable accuracy (e.g., \citealt{2003ApJ...582..905H}). 
A typical amount of $^{56}$Ni thus obtained for well-studied SNe is $\sim 0.07M_\odot$ (e.g., SN 1987A, SN 1994I, SN 2002ap; \citealt{1989ARA&A..27..629A}; \citealt{1994ApJ...437L.115I}; \citealt{2002ApJ...572L..61M}). 
Recently, the statistical analysis for more than 50 events of CCSNe has also suggested that the amount of $^{56}$Ni is around $0.07M\odot$ at the median \citep{2019MNRAS.485.1559P}.
Therefore, in `typical supernovae', on average $\sim0.07M_\odot$ of $^{56}$Ni should be synthesized.

Figure \ref{ fig:56Ni } shows the relation between the energy growth timescale $t_{\rm grow}$ and the synthesized $^{56}$Ni mass. 
The gray line represents the typical $^{56}$Ni mass in individual CCSNe, i.e., $0.07M_\odot$. 
Note that this figure adopts the deep ejected mass cut model, which is regarded to produce the maximally allowed amount of newly synthesized isotopes (see \S \ref{ sec:cut } \& \ref{ sec:mass-cut }).
It can be clearly seen that there is a decreasing tendency of $M({}^{56}{\rm Ni})$ for increasing $t_{\rm grow}$. 
For $M_{\rm ZAMS}=15M_\odot$, the models with $t_{\rm grow}= 10-100$ ms can reproduce the typical $^{56}$Ni mass.
For $M_{\rm ZAMS}=20M_\odot$, even the instantaneous explosion model ($t_{\rm grow}\sim10$ ms) does not produce the typical $^{56}$ Ni mass.
For $M_{\rm ZAMS}=25M_\odot$, $^{56}$Ni is synthesized abundantly, and the condition to produce $0.07M_\odot$ of $^{56}$Ni is satisfied with a wide range of  $t_{\rm grow}$ ($t_{\rm grow}= 10-500$ ms).
Since $\sim0.07M_\odot$ of $^{56}$Ni corresponds to the `average' mass ejected in various SNe, we can see that the explosion models of $t_{\rm grow}=10-500$ ms are consistent with the typical $^{56}$Ni mass.

The effect of different final/total energy on the amount of $^{56}$Ni is also shown in Figure \ref{ fig:56Ni }.
The amount of $^{56}$Ni is mainly determined by $t_{\rm grow}$, and is insensitive to $E_{\rm final}$.
Most importantly, since the synthesized amount of $^{56}$Ni roughly converges for $t_{\rm grow}\geqq100$ ms regardless of $E_{\rm final}$, it is understood that the synthesis of $^{56}$Ni is roughly determined by the explosion dynamics within 100ms after the initiation of the explosion.
This result is roughly consistent with the analytical estimates described in \S\ref{ sec:temp } (see also \citealt{2009MNRAS.394.1317M}).
Note that we use the model with $M_{\rm ZAMS}=20M_\odot$ and $E_{\rm final}=1.0\times10^{51}$ ergs for this discussion, but we have seen the same effect for $M_{\rm ZAMS}=$ 15$M_\odot$ and 25$M_\odot$.

The decreasing trend of $M({}^{56}{\rm Ni})$ can be understood as follows. 
$^{56}$Ni is mainly synthesized by the complete Si-burning at $T>5\times10^9$ K. 
As can be seen in Fig.\ref { fig:temp }, the peak temperature reached in the innermost region is decreased for increasing $t_{\rm grow}$. 
This is also confirmed by Figure \ref{ fig:abundance }. 
The argument presented for $M({}^{56}{\rm Ni})$ strongly supports the rapid explosion ($t_{\rm grow} < 500$ ms), and especially the best match is found for a nearly instantaneous explosion ($t_{\rm grow} \lesssim 100$ ms).
This result regarding the $^{56}$Ni production is consistent with the finding by \cite{2019MNRAS.483.3607S}, where further discussion is given especially on the connection of the $^{56}$Ni production with the initiation of the explosion.

We should also mention that the `neutrino-driven wind', which is the mass outflow from the PNS surface, may increase in the amount of $^{56}$Ni (\citealt{2018ApJ...852...40W}). 
However, since the amount of $^{56}$Ni from the neutrino-driven wind is at most $\lesssim0.005 M_\odot$ in the literature, this would not strongly affect our conclusions.
Consequently, in the slow explosion model ($t_{\rm grow}\geqq 1000$ ms), it is difficult to explain the broad range of $M(^{56}{\rm Ni})$ inferred for normal type II supernovae (0.005 to 0.28 $M_\odot$; \citealt{2017ApJ...841..127M}).
We conclude that the production of $^{56}$Ni is a strong constraint on the CCSN explosion mechanism; the slow explosion mechanism ($t_{\rm grow} \gtrsim 1000$ ms) would be inconsistent with observations for the typical CCSNe, for any progenitor masses and explosion energies.

\subsection{{\rm${}^{44}$Ti and ${}^{57}$Ni} produced in SN1987A}\label{ sec:87A }
Production of radioactive nuclei (such as $^{44}$Ti and $^{57}$Ni), which are assembled in the innermost region of an SN explosion, is sensitive to the condition in the innermost region of the exploding star, and is then sensitive to the dynamics of the SN explosion in the earliest phase. 
In addition to $^{56}$Ni, $^{44}$Ti and $^{57}$Ni can be detected directly or indirectly in some SNe and young SN remnants. 
This is the case for SN 1987A. 
The observational properties of SN 1987A have been extensively studied (e.g., \citealt{1988ApJ...330..218W}; \citealt{1989ARA&A..27..629A}; \citealt{1990ApJ...360..242S}; \citealt{1998ApJ...496..946K}; \citealt{1998ApJ...497..431K}; \citealt{2014ApJ...792...10S}). 
They provide observational estimates for the ejected masses of $^{57}$Ni and $^{44}$Ti. 
In this section, we discuss the appropriate timescale which must have been realized in the explosion of SN 1987A, as the best-studied CCSN, by comparing the synthesized masses of ${}^{44}$Ti and ${}^{57}$Ni in our models to the observational estimates. 
For the observational estimates, we adopt those provided by \cite{2014ApJ...792...10S}, 
which were obtained from a least square fit of the synthesized light curve (with different amounts of $^{57}$Ni and $^{44}$Ti) to the bolometric light curve of SN 1987A.
We note a caveat in our analysis; our simulation is not tuned to model SN 1987A. 
While our study covers a range of the initial masses ($15-25M_\odot$), the progenitor of SN1987A might evolve in a non-standard way ($\sim15-20M_\odot$; see, e.g., \citealt{1990ApJ...360..242S}; \citealt{2007AIPC..937..125P}). 
Therefore, while this would not strongly affect our conclusions, the constraint in this section should be regarded as being qualitative.

\begin{figure}
\begin{center}
  \includegraphics[width=80mm]{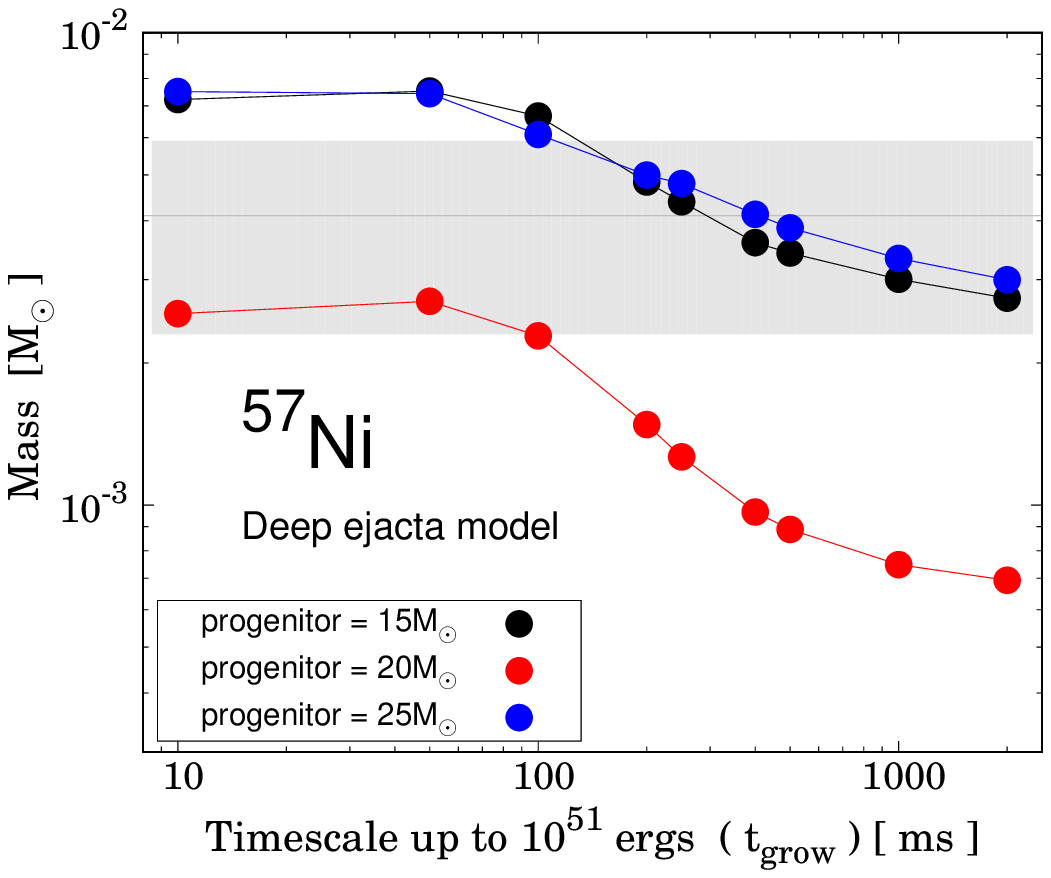}
  \includegraphics[width=80mm]{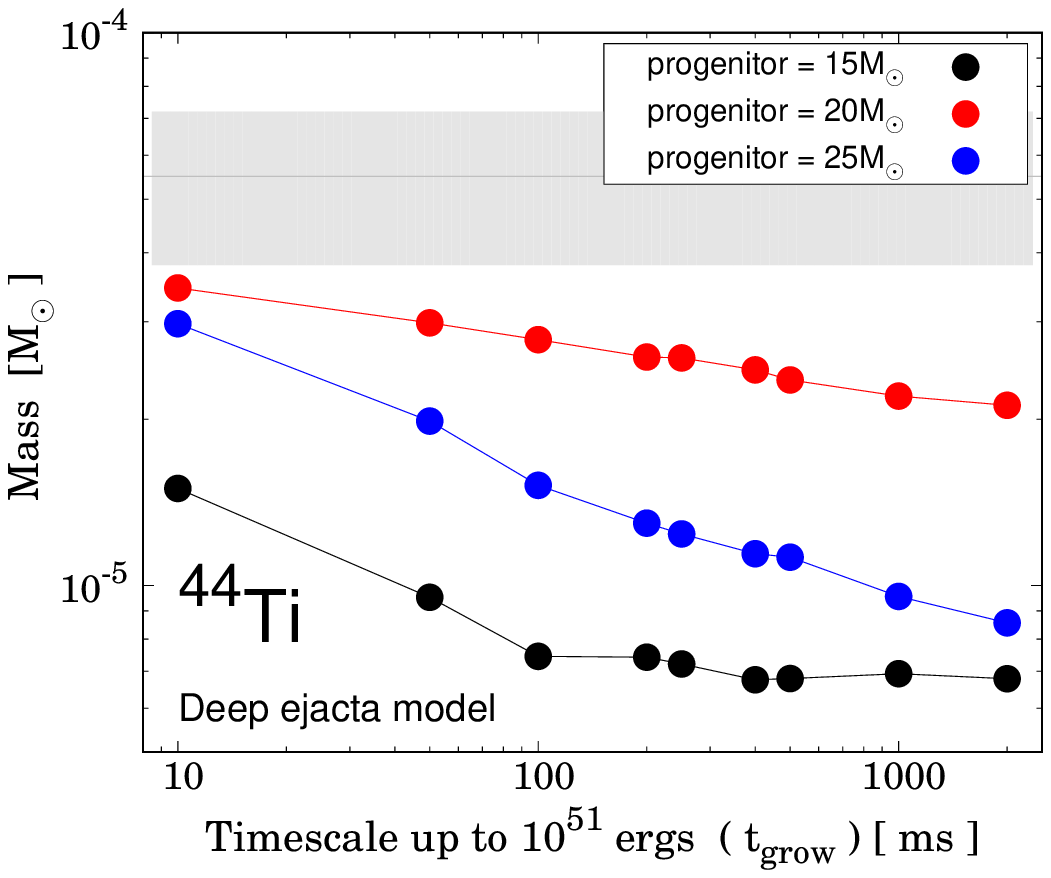}
  \caption{The energy growth timescale $t_{\rm grow}$ and the produced $^{57}$Ni and $^{44}$Ti mass. 
Show here is dependence on $M_{\rm ZAMS}$ (for given $E_{\rm final}=1.0\times10^{51}$). 
The gray region shows $M({\rm {}^{57}Ni}) =0.0041\pm 0.0018M_\odot$ and $M({\rm {}^{44}Ti}) =0.55\pm 0.17 \times10^{-4}M_\odot$, corresponding to the observational estimate for SN1987A taken from \cite{2014ApJ...792...10S}. }
  \label{ fig:57Ni } 
\end{center}
\end{figure}
\begin{figure}
\begin{center}
  \includegraphics[width=80mm]{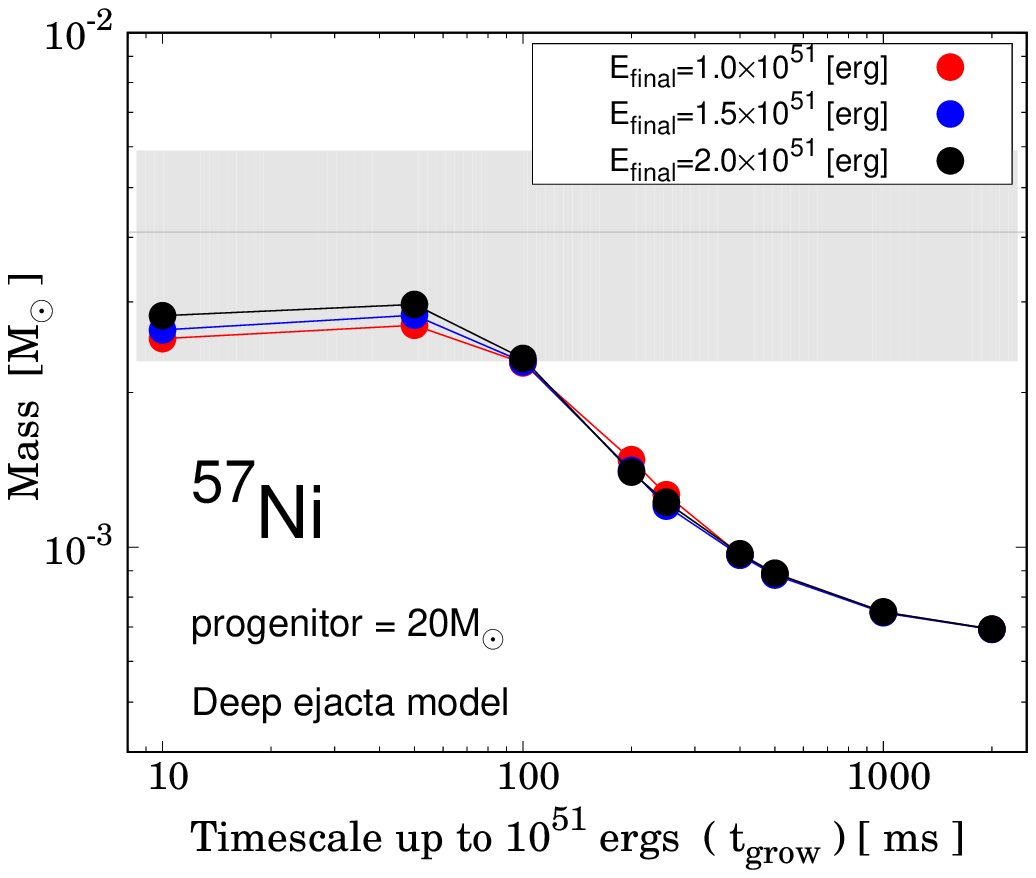}
  \includegraphics[width=80mm]{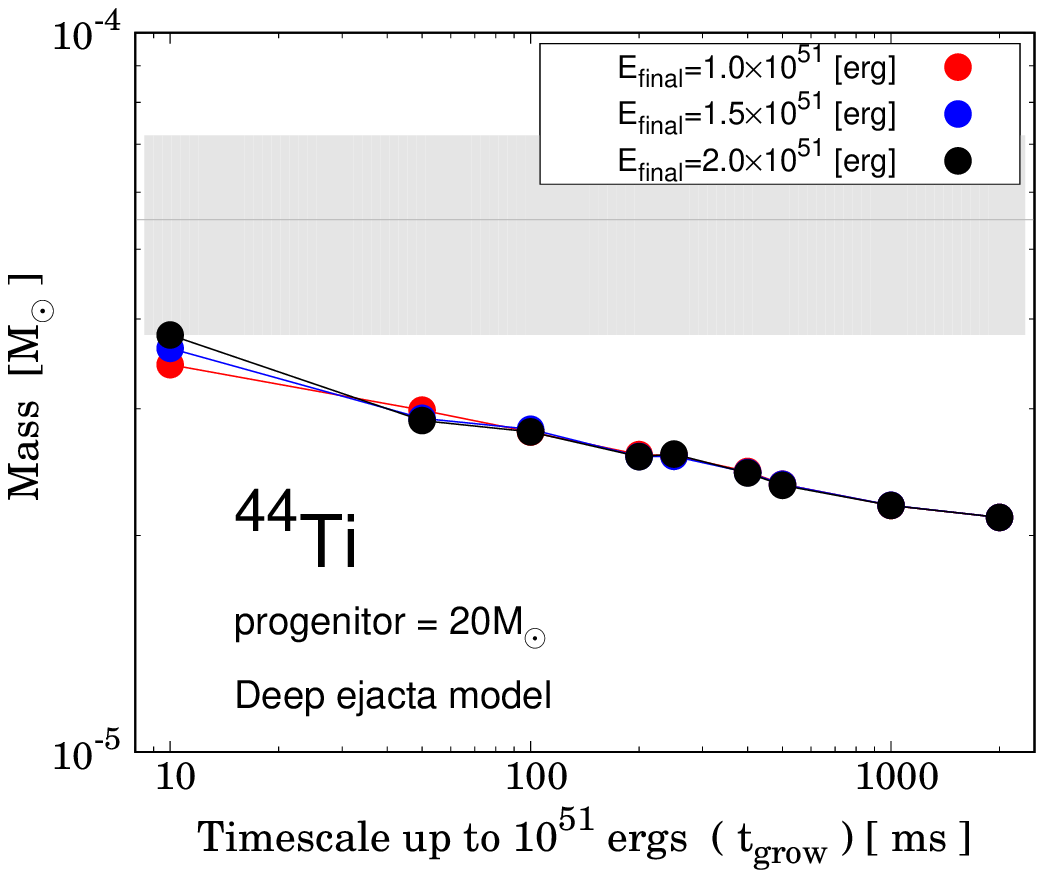}
  \caption{Same as Fig.\ref{ fig:57Ni }, but for dependence on $E_{\rm final}$ (for given $M_{\rm ZAMS}=20M_\odot$).}
  \label{ fig:44Ti } 
\end{center}
\end{figure}

Figures \ref{ fig:57Ni } and \ref{ fig:44Ti } show the relation between the energy growth timescale $t_{\rm grow}$ and the synthesized $^{57}$Ni and $^{44}$Ti masses. 
The gray regions in the figures correspond to the observational estimate ($M({\rm {}^{57}Ni}) =0.0041\pm 0.0018M_\odot$, and $M({\rm {}^{44}Ti}) =0.55\pm 0.17 \times10^{-4}M_\odot$; \citealt{2014ApJ...792...10S}). 
For both of $^{44}$Ti and $^{57}$Ni, we find there is a clear decreasing trend for increasing $t_{\rm grow}$ similarly to the case of $^{56}$Ni. 
Furthermore, the synthesized mass deviates from the observational estimates of SN 1987A, as $t_{\rm grow}$ increases. 
Fig \ref{ fig:57Ni } shows that this tendency does not depend on $M_{\rm ZAMS}$. 
Moreover, the masses of $^{57}$Ni and $^{44}$Ti are determined by $t_{\rm grow}$ regardless of $E_{\rm final}$, as shown in Figure \ref{ fig:44Ti }.
Figure \ref{ fig:44Ti } shows that the synthesized amount of $^{57}$Ni ($^{44}$Ti) converges for $t_{\rm grow}\geqq100$ ms.
This result is attributed to the decrease in the peak temperature in the central region (Figs.\ref { fig:temp } \& \ref{ fig:abundance }), following the same discussion for $^{56}$Ni. 
The iron group elements are mainly produced through the complete/incomplete Si-burning, which takes place in the innermost region with a very high temperature.

We should mention that our model can only marginally explain the mass of $^{44}$Ti even for the best case with small $t_{\rm grow}$.
As mentioned above, the result may be influenced by $M_{\rm ZAMS}$, and the influence is not monotonous to $M_{\rm ZAMS}$ (see Figure \ref{ fig:57Ni }). 
Therefore, $M_{\rm ZAMS}$, which is optimal for reproducing the observation of SN1987A, may exist in the parameter not adopted by this study.
Furthermore, multi-dimensional effects in the explosion may also be a key in the synthesis of $^{44}$Ti; for example, a jet-like explosion is suggested to increase the amount of $^{44}$Ti (\citealt{1998ApJ...492L..45N}; \citealt{2003ApJ...598.1163M}). 
However, these effects would not be sufficiently strong to remedy the large discrepancy we find here. 
Indeed, the jet-like explosion can lead to the high ratio of $^{44}$Ti/$^{56}$Ni, but the mass of $^{44}$Ti itself tends to decrease for a decreasing amount of $^{56}$Ni (\citealt{2003ApJ...598.1163M})\footnote{Another caveat on $^{44}$Ti is possible production within the neutrino-driven wind \citep{2017ApJ...842...13W}. However, the masses of $^{56}$Ti and $^{57}$Ti would not be significantly affected by this process.}. 
Consequently, a combined analysis of $^{57}$Ni, $^{44}$Ti and $^{56}$Ni is taken as a strong constraint; the slow explosion mechanism (especially $t_{\rm grow} \gtrsim 1000$ ms) is disfavored to explain the properties of SN 1987A.

\subsection{Comparison to the abundances of the extremely metal-poor stars}\label{ sec:MPS }

In the early phase of the galaxy formation when the metallicity was still low, the entire galaxy was not yet chemically mixed substantially. 
Therefore, a local metal abundance of the gas is believed to be represented by a single supernova event (e.g., \citealt{1995ApJ...451L..49A}). 
Low mass stars formed in such an early phase of the galaxy formation survive until today, and they are observed as extremely metal-poor (EMP) stars. 
We can, therefore, assume that the EMP stars preserve abundance patterns of individual CCSNe in the early Universe. 
This strategy has been adopted to constrain the nature of population III or II massive stars and their SN explosions (e.g., \citealt{2007ApJ...660..516T}; \citealt{2010ApJ...724..341H}). 
We follow this approach, by requiring that the typical CCSN yields should be consistent with the abundance patterns of the EMP stars, and consider to constrain the explosion mechanism of `typical supernovae'. 

In this section, by comparing our results to [X/Fe] observed in the EMP stars (i.e., [Fe/H]$\lesssim-2.5$), we aim at constraining an appropriate range of $t_{\rm grow}$. 
Note that our model assumes the solar metallicity. 
Therefore, here we focus on the elements produced via explosive burning, which tend to be insensitive to the progenitor metallicity (see Figure 1-5 in \citealt{2006ApJ...653.1145K}), noting that we omit the metal content already having existed at ZAMS (see section \ref{ sec:mass-cut }). 
Despite the caveat, this is a useful exercise; 
if the range of $t_{\rm grow}$ would turn out to be completely inconsistent with the EMP star constraints, 
even with the additional parameters ($M_{\rm ZAMS}$, $E_{\rm final}$ and metallicity $Z_{\rm ZAMS}$), 
such a model should still experience a major difficulty in reproducing the abundances of the EMP stars.

Here, we consider the elements from Mn to Zn.
As discussed in the later section \ref{ sec:discussion }, the composition of the light elements (Ti, Cr) turn out to be relatively insensitive to $t_{\rm grow}$, compared to other parameters. 
This section focus on Mn, Co, and Zn which are sensitive to $t_{\rm grow}$.
Figures \ref{ fig:Zn }-\ref{ fig:Co } show the heavy-element abundance ratios [X/Fe] as a function of the energy growth timescale $t_{\rm grow}$. 
The instantaneous explosion model ($t_{\rm grow}\lesssim 100$ ms) reproduces observational data fairly well. 
On the contrary, the slow explosion model ($t_{\rm grow}\gtrsim 400$ ms) results in the abundance ratios totally different from those observed in the EMP stars. 
In the following subsections, we discuss the details for each element.

\subsubsection{ Zn }\label{ sec:Zn }
\begin{figure}
\begin{center}
  \includegraphics[width=80mm]{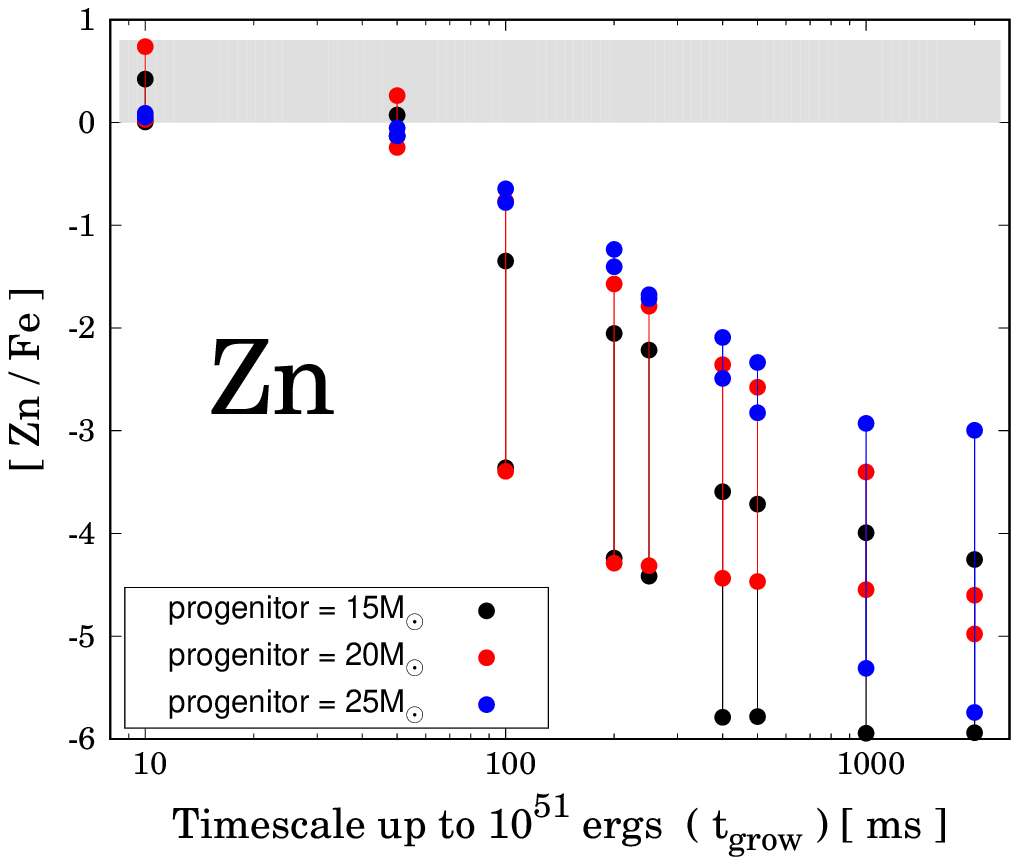}
  \includegraphics[width=80mm]{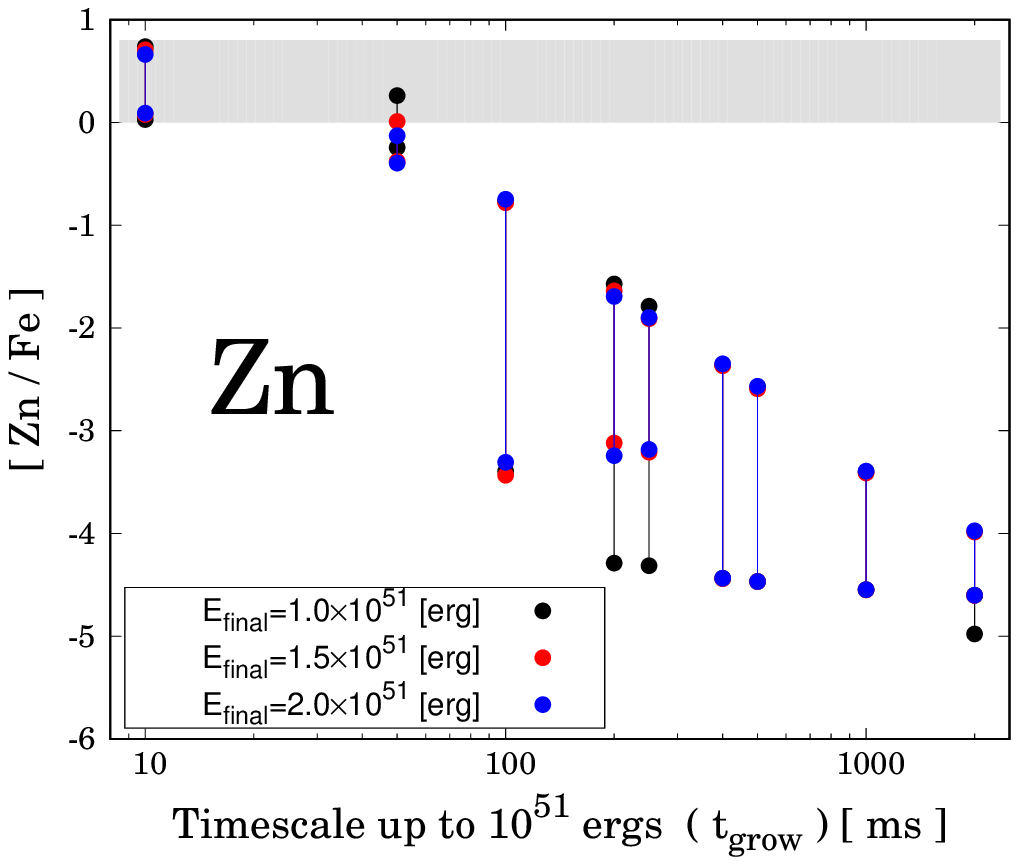}
  \caption{The relation between the energy growth timescale and [Zn/Fe]. 
The left panel shows the models with $M_{\rm ZAMS}=$ 15, 20, 25 $M_\odot$, all of which have $E_{\rm final}=1.0\times10^{51}$ erg. 
The right panel is for $E_{\rm final}=$ 1.0, 1.5, 2.0$\times10^{51}$ erg with $M_{\rm ZAMS}=$ 20$M_\odot$. 
The gray region corresponds to $-0.2<$[Zn/Fe]$<0.6$, which is the range observed in the EMP stars (\citealt{1995AJ....109.2757M}; \citealt{1996ApJ...471..254R}; \citealt{2004A&A...416.1117C}; \citealt{2004ApJ...607..474H}).
For each model, there are two model yields as shown by the filled circles connected by a line with the same color, for the two choices of the mass cut (see section.\ref{ sec:cut }). We regard this range as a model uncertainty in this analysis. }
  \label{ fig:Zn } 
\end{center}
\end{figure}
Figure \ref{ fig:Zn } shows the relation between the energy growth timescale and [Zn/Fe]. 
The gray region in the figure corresponds to $-0.2<$[Zn/Fe]$<0.6$, which is the range observed for the EMP stars (\citealt{1995AJ....109.2757M}; \citealt{1996ApJ...471..254R}; \citealt{2004A&A...416.1117C}; \citealt{2004ApJ...607..474H}). 
The error bar corresponds to the uncertainty due to the amount of fallback, i.e. the mass cut position.
The instantaneous explosion model ($t_{\rm grow}$= 10-50 ms) is consistent with [Zn/Fe] observed for the EMP stars. 
We can see that the production of Zn is very sensitive to $t_{\rm grow}$. 
If $t_{\rm grow} \geq 400$ ms, there is virtually no production of Zn, resulting in [Zn/Fe] $\lesssim -3$ and thus a large discrepancy from the observations.
While we have tested different values of $M_{\rm ZAMS}$ and the mass cut position, as seen in Figure \ref{ fig:Zn }, this large difference is not remedied by changing these parameters.
In addition, within our formalism, $E_{\rm final}$ does not affect the value of [Zn/Fe] as also shown in Figure \ref{ fig:Zn }.

Dependence of [Zn/Fe] on the electron fraction Ye (that is, implicitly the metallicity of the progenitor) may require further consideration.  
For larger Ye ($\sim0.5$), the amount of $^{64}$Ge ($\to$ $^{64}$Zn) will be enhanced, and likewise $^{56}$Ni ($\to$ $^{56}$Fe) production will be enhanced. 
Therefore, the resulting ratio of Zn/Fe depends to some extent on Ye but not strongly.
\cite{2006ApJ...653.1145K} indicate that this effect of the initial metallicity on [Zn/Fe] is within $\pm0.2$. 
Thus, this discrepancy would not be remedied by simply changing the metallicity of the progenitor.

This tendency can be understood as follows. 
In the model, Zn is dominated by $^{64}$Zn as a decay product of $^{64}$Ge. 
This isotope is mainly produced through the $\alpha$-rich freeze-out, which takes place in the innermost region with the very high temperature. 
Therefore the production of Zn is suppressed for the slow explosion. 
Consequently, we conclude that Zn/Fe is an important indicator of the explosion mechanism, as it responds sensitively to the nature of the explosion in the central region.


\subsubsection{ Co and Mn }\label{ sec:Co }
\begin{figure}
\begin{center}
  \includegraphics[width=80mm]{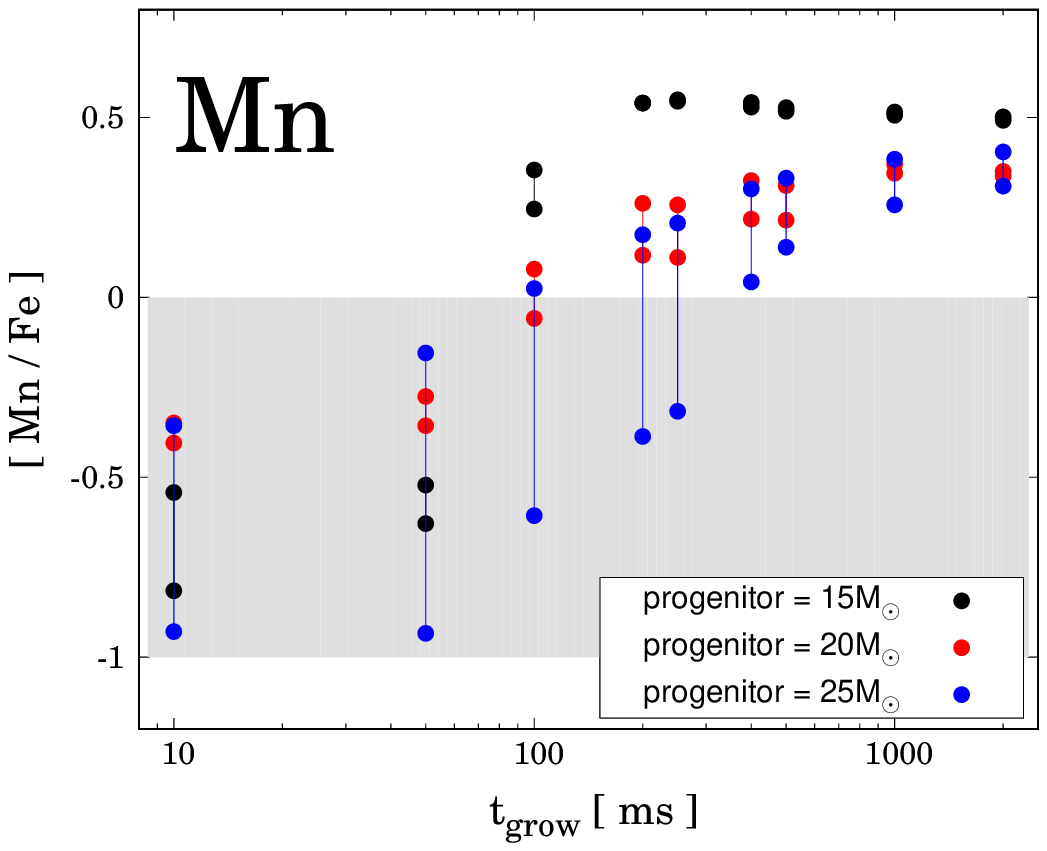}
  \includegraphics[width=80mm]{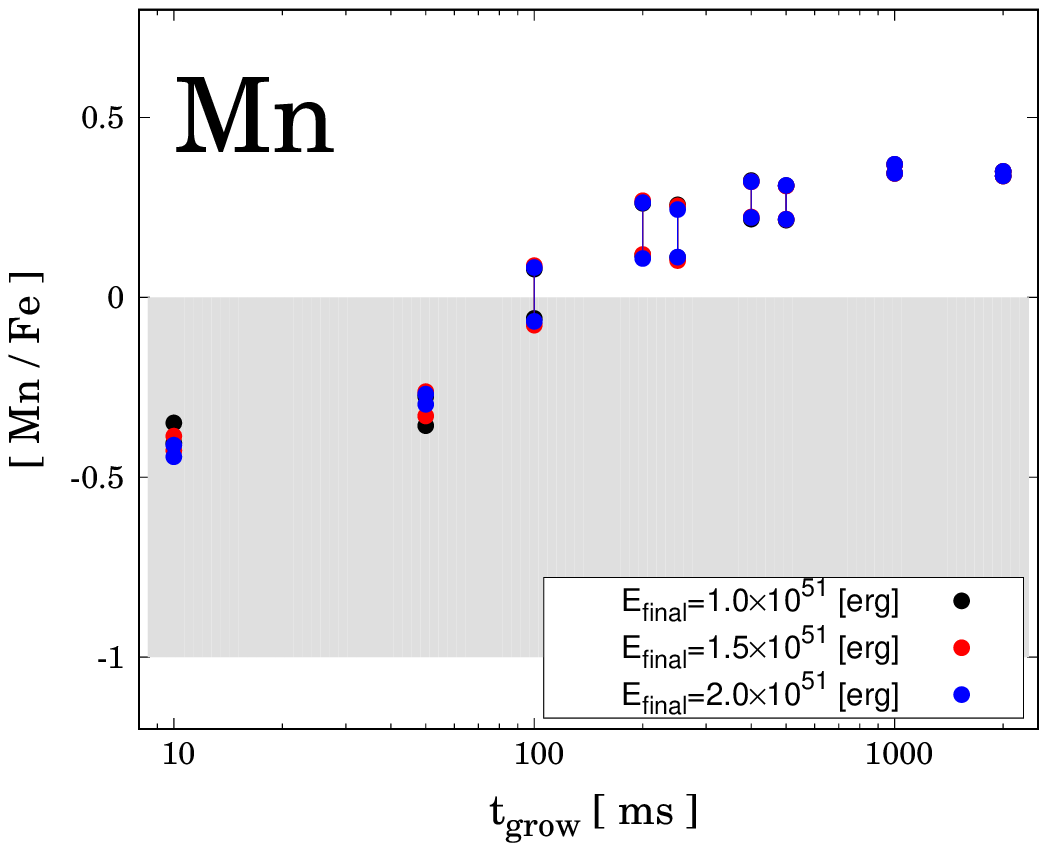}
  \caption{Same as Fig.\ref{ fig:Zn }, but for [Mn/Fe] .The gray region corresponds to $-1.0<$[Mn/Fe]$<0.0$. }
  \label{ fig:Mn } 
\end{center}
\end{figure}
\begin{figure}
\begin{center}
  \includegraphics[width=80mm]{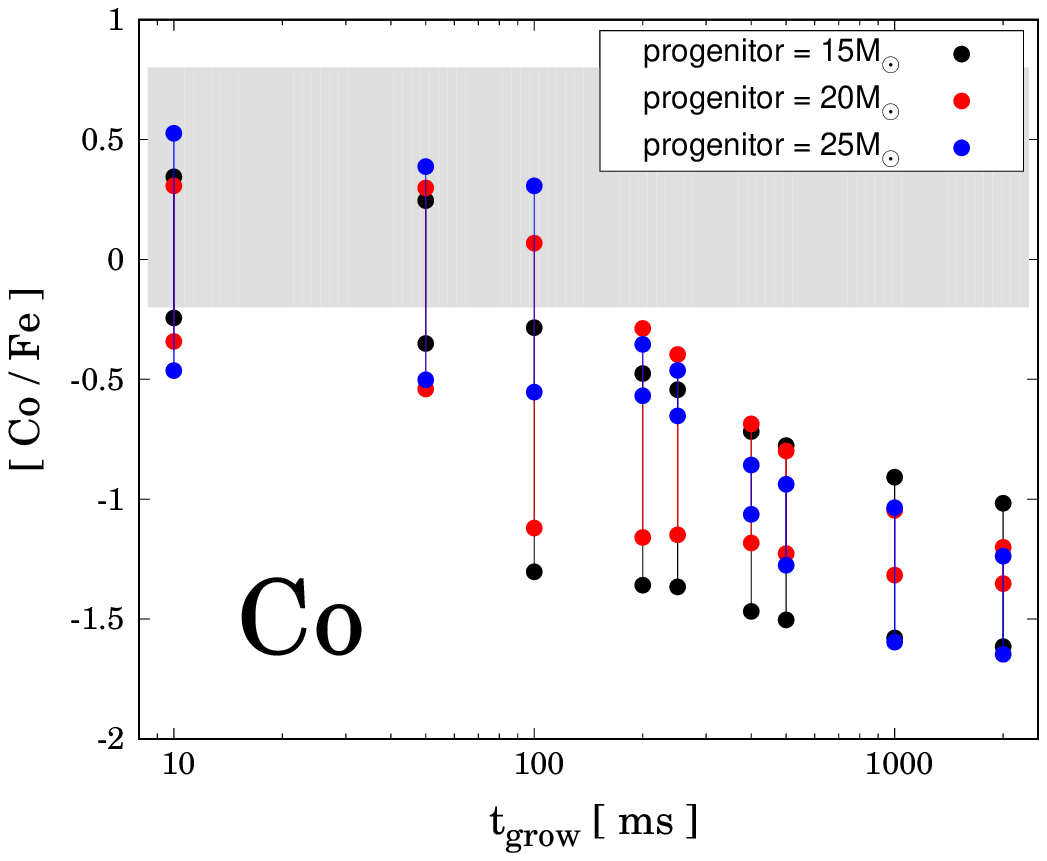}
  \includegraphics[width=80mm]{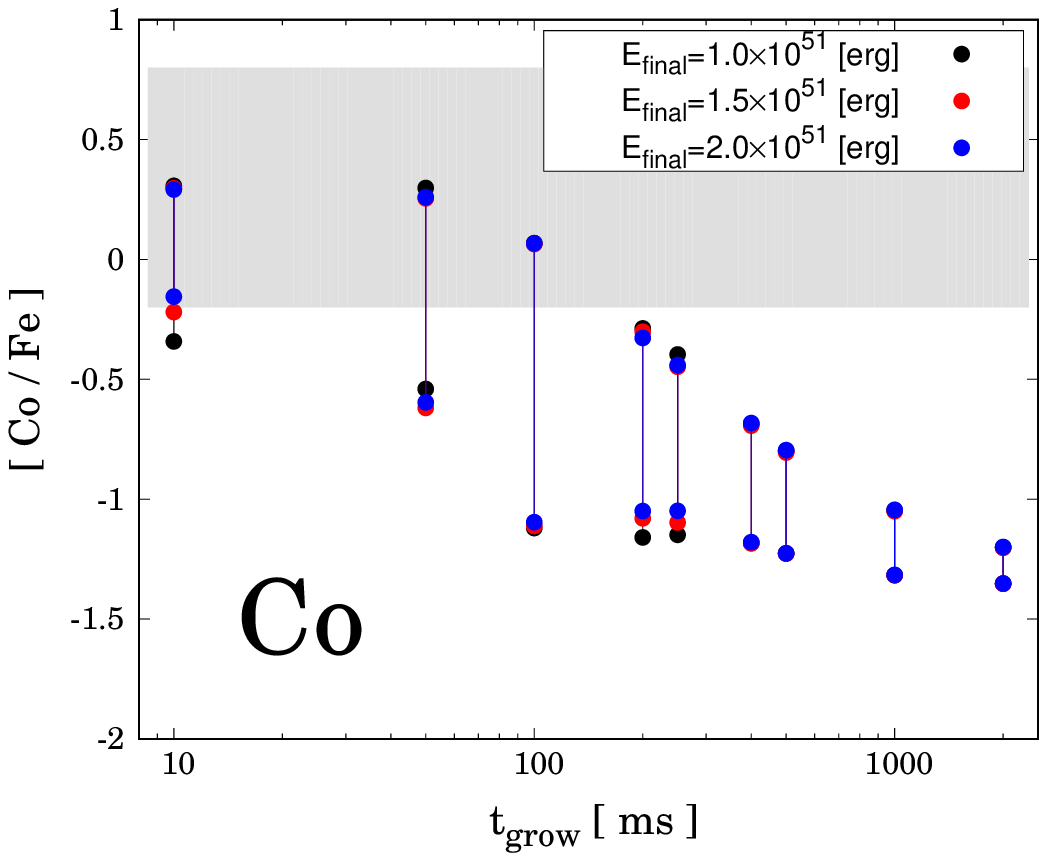}
  \caption{Same as Fig.\ref{ fig:Zn }, but for [Co/Fe]. The gray region corresponds to $-0.2<$[Co/Fe]$<0.8$. }
  \label{ fig:Co } 
\end{center}
\end{figure}

The relation between $t_{\rm grow}$ and [Mn/Fe] is shown in Figure \ref{ fig:Mn }. The same for [Co/Fe] is shown in Figure \ref{ fig:Co }. 
The ranges observed for the EMP stars are $-1.0<$[Mn/Fe]$<0.0$ and $-0.2<$[Co/Fe]$<0.8$, respectively. 
These figures clearly show that the productions of both Mn and Co are very sensitive to $t_{\rm grow}$. 
The results from the rapid explosion model ($t_{\rm grow} < 250$ ms) are again consistent with the observational ratios of [Mn/Fe] and [Co/Fe]. 
Moreover, there is the increasing trend of [Mn/Fe] and the decreasing trend of [Co/Fe] for increasing $t_{\rm grow}$, resulting in a substantial discrepancy between the observations and the slow explosion models ($t_{\rm grow} \gtrsim 1000$ ms). 
Specifically, the slow explosion models result in [Mn/Fe] $\sim 0.5$ and [Co/Fe] $\sim -1$, failing to explain the observed ratios of [Mn/Fe] $\sim -0.5$ and [Co/Fe] $\sim 0.2$.

These elements, Mn and Co, are dominated by the decay products of $^{55}$Co and $^{59}$Cu, respectively. 
$^{59}$Cu is an isotope with neutron excess, and synthesized mainly by the complete Si-burning in the deepest layer of the ejecta. 
Following our discussion on the extent of different burning layers (\S \ref{ sec:abundance }), the production of Co (and, thus, [Co/Fe]) is suppressed for the slow explosion model. 
In contrast, $^{55}$Co is also an isotope with neutron excess but formed through the incomplete Si-burning in the outer layer. 
[Mn/Fe] is increased for increasing $t_{\rm grow}$, due to the suppression in the production of $^{56}$Ni.  
Despite the ambiguity in $M_{\rm ZAMS}$ and $E_{\rm final}$, this general trend will not be altered significantly by adopting a range of these parameters, as seen in Figures \ref{ fig:Mn } and \ref{ fig:Co }. 
Regarding the ambiguity in Ye (that is, implicitly the metallicity of the progenitor), both of $^{55}$Co and $^{59}$Cu are neutron rich isotopes, so both of [Mn/Fe] and [Co/Fe] will tend to be smaller for larger Ye. 
Therefore, a better match to both isotopes by changing Ye would never be obtained.
These opposite trends strongly support the rapid explosion model ($t_{\rm grow} \lesssim 250$ ms). 
In summary, we conclude that Mn/Fe and Co/Fe are important indicators which reflect the nature of the explosion regardless of the natures of the progenitor stars (i.e., mass $M_{\rm ZAMS}$ and metallicity $Z_{\rm ZAMS}$).

\section{Summary and Discussion}\label{ sec:discussion }

In this paper, we have studied the CCSN explosive nucleosynthesis by parameterizing the nature of the explosion mechanism by the timescale $t_{\rm grow}$.
$t_{\rm grow}$ is defined as the energy growth timescale in which the explosion energy is reached to $10^{51}$ ergs since the initiation of the explosion. 
By using 1D-hydrodynamics and nucleosynthesis calculations, we have shown how the nucleosynthesis products are affected by $t_{\rm grow}$. 
We have then compared our numerical results to various observational constraints; 
the masses of $^{56}$Ni derived for typical CCSNe, 
the masses of $^{57}$Ni and $^{44}$Ti observed for SN 1987A, 
and the abundance patterns observed in extremely metal-poor stars. 
We find that these observational constraints are consistent with the `rapid' explosion ($t_{\rm grow} < 250$ msec), 
and especially the best match is found for a nearly instantaneous explosion ($t_{\rm grow} \lesssim 50$ msec). 
The discrepancy is larger for larger $t_{\rm grow}$.
Indeed, the synthesis of Fe-group elements is roughly determined by the explosion dynamics within 100 msec after the initiation of the explosion.
Even if we take into account the uncertainties in the natures of the progenitor (e.g., ZAMS
masses $M_{\rm ZAMS}$ and metallicity $Z_{\rm ZAMS}$), it is very unlikely that the slow explosion model ($t_{\rm grow} \gtrsim 1000$ msec) can satisfy all these constraints. 
Therefore, as a robust conclusion, the slow explosion mechanism ($t_{\rm grow} \gtrsim 1000$ msec), which is seen in recent ab-initio simulations, is strongly disfavored as an explosion mechanism of typical CCSNe (Figure \ref{ fig:summary }).

The final energy of the explosion ($E_{\rm final}$) has been adopted from the canonical value ($E_{\rm ref} \equiv 10^{51}$ ergs) to twice the value in this paper. 
Observationally, there is a broad range of $E_{\rm final}$ in CCSNe. 
Note that $t_{\rm grow}$ is defined in a way so that it is independent from $E_{\rm final}$; 
this is the timescale in which the explosion energy is reached to $E_{\rm ref} \equiv 10^{51}$ ergs since the initiation of the explosion. 
Namely, $t_{\rm grow}$ can be translated to the energy deposition rate $\dot E_{\rm exp}$ through a one-to-one relation regardless of $E_{\rm final}$. 
For each progenitor model with the binding energy $E_{\rm bind}$, the energy deposition rate $\dot E_{\rm exp}$ is therefore obtained as  $\dot E_{\rm exp}\equiv (E_{\rm ref}+|E_{\rm bind}|)/t_{\rm grow}$. 
If the timescale of the explosion is defined this way, then the slow model ($t_{\rm grow} \gtrsim 1000$ msec) for $M_{\rm ZAMS}=20M_\odot$ represents $\dot E_{\rm exp} < 1.31 \times 10^{51}$ erg s$^{-1}$, and our analysis rejects such models to represent the typical CCSNe explosion mechanism.

We emphasize that the amount of Fe-peak elements produced by our models converges for $t_{\rm grow}\geqq100$ msec. 
Therefore, it is understood that the synthesis of $^{56}$Ni (and generally the complete Si burning products) is roughly determined in the explosion dynamics within 100msec after the initiation of the explosion.
Namely, the need of the `rapid' explosion means that the explosion mechanism  should realize the rapid increase of the explosion energy  within the first 100 msec (Figure \ref{ fig:summary }); the later evolution is indeed not important in terms of the Fe-peak element synthesis. 
This result is consistent with the analytical estimate described by \cite{2009MNRAS.394.1317M}). 
The further discussed the dependence of $M({}^{56}{\rm Ni})$ on $E_{\rm exp}$ and $\dot{E}_{\rm exp}$ analytically. 
For a given $E_{\rm exp}$, $M({}^{56}{\rm Ni})$ is saturated above a certain critical value of $\dot{E}_{\rm exp}$. 
Below the critical value ($\dot{E}_{\rm cr}$), $M({}^{56}{\rm Ni})$ is smaller for smaller  $\dot{E}_{\rm exp}$.
This critical value here ($\dot{E}_{\rm cr}$) is given by $\dot{E}_{\rm cr,51}(E_{\rm exp})=(2.9/T_9)^{-10/3}\rho_6^{-1/2}E_{\rm exp,51}^{2/3}$, where $\dot{E}_{\rm cr,51}=\dot{E}_{\rm cr}/10^{51}{\rm erg/s}$,  $E_{\rm exp,51}=E_{\rm exp}/10^{51}{\rm erg}$, $T_9=T/10^9{\rm K}$ and $\rho_6=\rho/10^6{\rm g\cdot cm^{-3}}$. 
By noting that $\dot{E}(t)= E(t)/t$, where $E(t)$ is the energy released by time $t$, we can draw a dividing line in the $t-E$ plain which characterize the evolution of the explosion energy growth (Figure \ref{ fig:summary }); below this line, the production of $^{56}$Ni (and Fe-peaks) is suppressed as compared to the instantaneous explosion. This line is defined as $E_{\rm exp}=(2.9/T_9)^{-10}\rho_6^{-3/2}\cdot t^3$.
The present conclusion can then be rephrased as follows; the energy growth must be sufficiently rapid so that the production of $^{56}$Ni (and Fe-peaks) is saturated to the value expected in the instantaneous explosion. 
Otherwise, the production of these elements/isotopes is suppressed as compared to the instantaneous explosion, leading to the discrepancies to various observations. 
Therefore, in testing the validity of the ab-initio simulations in view of the nucleosynthesis yields, a focus must be placed on the evolution of the energy in the first $\sim$ 100 msec after the initiation of the explosion (Figure \ref{ fig:summary }).

\begin{figure}
\begin{center}
  \includegraphics[width=80mm]{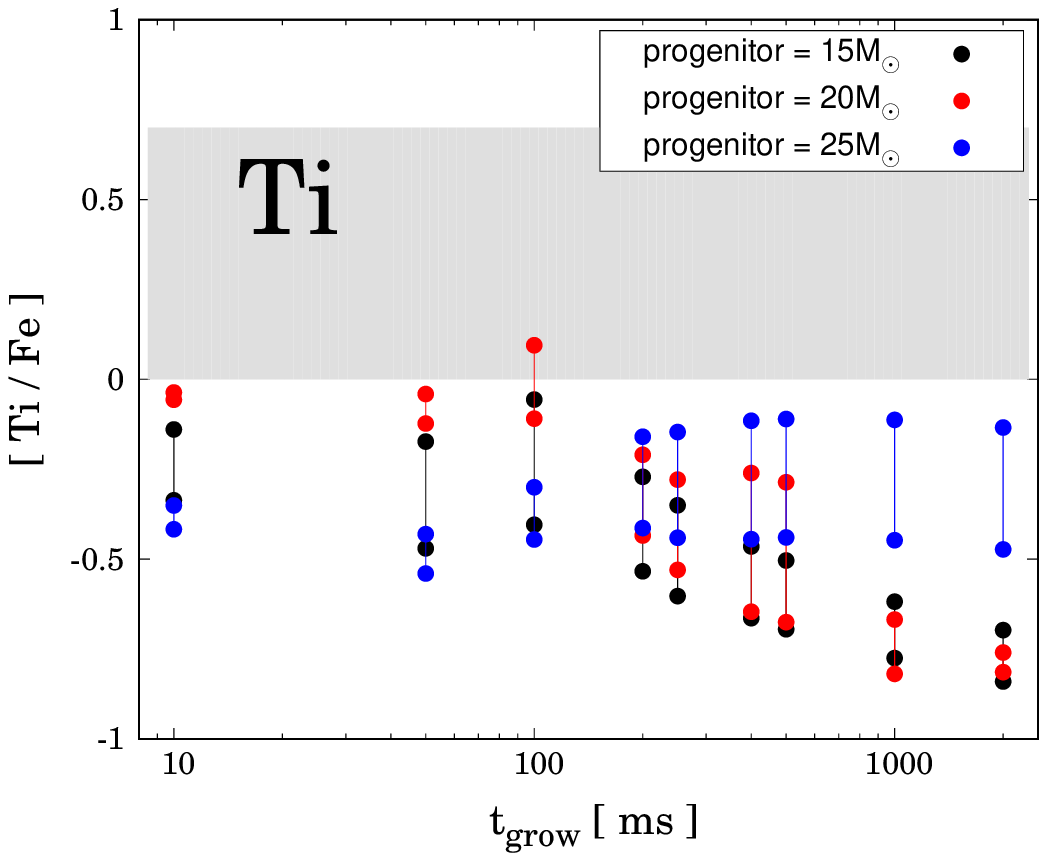}
  \includegraphics[width=80mm]{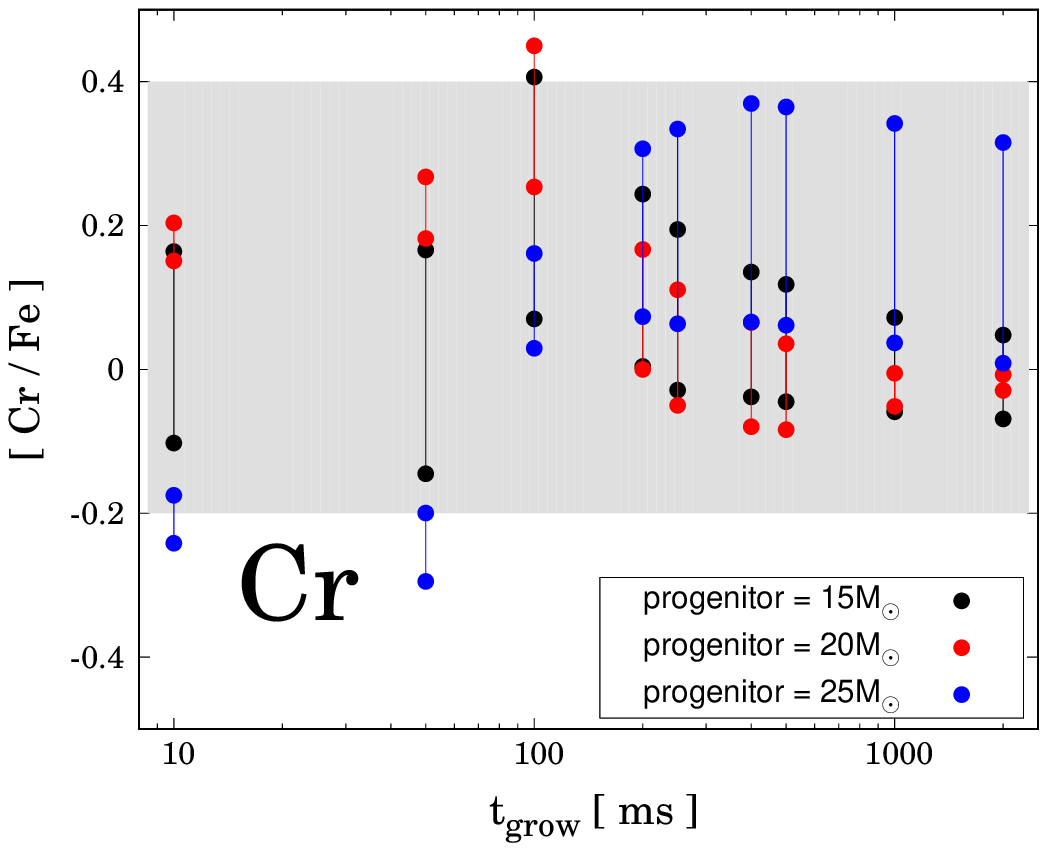}
  \caption{Same as Fig.\ref{ fig:Zn }, but for $t_{\rm grow}$ and [Ti/Fe], [Cr/Fe].The gray regions correspond to $0.0<$[Ti/Fe]$<0.7$ and $-0.2<$[Cr/Fe]$<0.4$. }
  \label{ fig:TiCr } 
\end{center}
\end{figure}

\begin{figure}[htb]
\begin{center}
  \includegraphics[width=80mm]{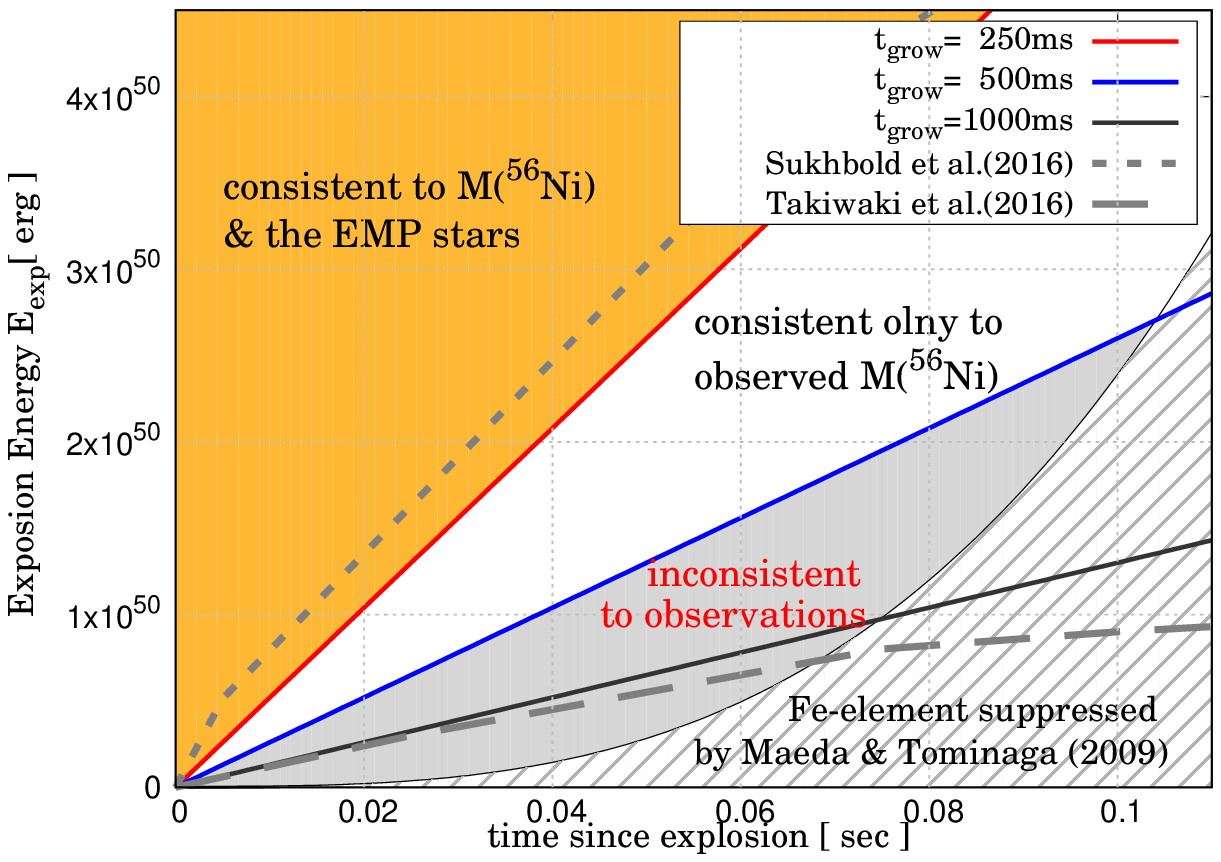}
  \caption{Summary of our results in terms of the time evolution of the explosion energy.
The x-axis corresponds to time after the initiation of the explosion, and y-axis to the explosion energy $E_{\rm exp}$ generated by that moment. 
The gray dotted and dashed lines show the time evolution of the explosion energy by \cite{2016ApJ...821...38S} (`artificial' explosion) and that by \cite{2016MNRAS.461L.112T} (ab-initio explosion), respectively. 
The hatched region shows the region where the production of the Fe-group elements is suppressed as compared to the instantaneous explosion, following an analytic treatment by  \cite{2009MNRAS.394.1317M}. 
The slow model ($t_{\rm grow} \geq 1000$ msec), which corresponds to $\dot E_{\rm exp} \leqq1.3\times 10^{51}$ erg s$^{-1}$ in our formalism, is rejected as the typical CCSNe explosion mechanism. }
  \label{ fig:summary } 
\end{center}
\end{figure}

To further expand the discussion in \S \ref{ sec:MPS } for the abundances of the EMP stars, here we discuss the abundance of Ti and Cr. 
The lighter elements from O to Ca are not discussed here;
these elements are mainly included in the outer layer, 
and thus the abundance ratio of (some of) these elementals to iron can be highly dependent not only on their fraction but also the iron fraction at ZAMS (i.e., metallicity). 
The dependences of [Ti/Fe] and [Cr/Fe] on $t_{\rm grow}$ are shown in Figure \ref{ fig:TiCr }.  
For both Ti and Cr, the ratios are relatively insensitive to $t_{\rm grow}$, and they are affected more substantially by the progenitor mass $M_{\rm ZAMS}$ and the mass cut position.
Here, Figure \ref{ fig:TiCr } shows that the [Ti/Fe] realized in our model is below the observational constrain, which cannot be improved by changing our parameters ($t_{\rm grow}$, $E_{\rm final}$, $M_{\rm ZAMS}$, and mass cut position). 
As described in the discussion for $^{44}$Ti in \S \ref{ sec:87A }, this may be remedied by a jetlike explosion with high entropy (\citealt{1998ApJ...492L..45N}; \citealt{2003ApJ...598.1163M}).
This is indeed a problem generally seen Galactic chemical evolution study \citep{2006ApJ...653.1145K}.
It is anyway concluded that the ratio of Ti and Cr to iron is insensitive to the nature of the explosion, especially to the explosion timescale $t_{\rm grow}$.
In summary, we conclude that considering these ratios would not alter our conclusions.
Indeed, we emphasize that these ratios are inappropriate to be an indicator of the explosion mechanism.

Our findings are summarized as follows: 

${\bf1.~~}$
The amount of synthesized $^{56}$Ni serves as a strong constraint on the CCSN explosion mechanism
(which confirms the suggestion by \citealt{2019MNRAS.483.3607S}); 
the slow explosion mechanism (that is, an increase of $t_{\rm grow}$) tends to suppress the production of $^{56}$Ni and would not satisfy various observational constraints, even taking into account the main uncertainties, such as the mass-cut position, $M_{\rm ZAMS}$ and $E_{\rm final}$. 
Especially, the argument related to $M({}^{56}{\rm Ni})$ strongly supports the rapid explosion ($t_{\rm grow} < 500$ msec), and the models with $t_{\rm grow}\gtrsim1000$ msec would never explain the nature of typical CCSNe. 
The amount of synthesized $^{56}$Ni is important for diagnosing the explosion timescale $t_{\rm grow}$.

${\bf2.~~}$
In view of the observational properties of SN1987A, we have found that the masses of synthesized $^{57}$Ni and $^{44}$Ti provide strong constraints to the CCSN explosion mechanism; 
the instantaneous explosion model ($t_{\rm grow} \lesssim 100$ msec) can roughly satisfy these constraints, 
while the increase of $t_{\rm grow}$ tends to suppress the synthesized amounts of these isotopes.
Note that this simulation is not tuned to mimic SN1987A and multi-dimensional effects in the explosion may also be a key in the synthesis of $^{44}$Ti (\citealt{1998ApJ...492L..45N}; \citealt{2003ApJ...598.1163M}). 
However, we find that combined analysis of $^{56}$Ni, $^{57}$Ni, and $^{44}$Ti strongly disfavors the slow explosion model with $t_{\rm grow} \gtrsim 1000$ msec as the CCSNe explosion mechanism.

${\bf3.~~}$ 
By comparing the model yields to [Zn, Co, Mn/Fe] of the EMP stars, we have found that the discrepancy from the observational properties is more significant for larger $t_{\rm grow}$. 
These abundance patterns in the EMP stars strongly support the instantaneous explosion model ($t_{\rm grow} \lesssim 50$ msec); the slow explosion mechanism (especially $t_{\rm{grow}}\gtrsim$1000 msec)  is unable to explain the chemical enrichment in the Galaxy. 
While we have tested different values of $M_{\rm ZAMS}$, $E_{\rm final}$ and the mass cut position, this large difference could not be remedied by changing these parameters.
Indeed, the different trends (either increasing or decreasing) as a function of $t_{\rm grow}$ for some elements' ratios (e.g., [Mn/Fe] vs. [Co/Fe]) as a function of $t_{\rm grow}$ make it nearly impossible to remedy the discrepancy found for the slow explosion models by changing these parameters.
Consequently, we suggest that Zn/Fe, Mn/Fe, and Co/Fe are important indicators for the nature of the explosion.
In addition, we also suggest that the ratio of either Ti or Cr to Fe would not be a good indicator for the explosion mechanism.

While our analysis is based on various simplifications following the classical thermal bomb-type simulations, we believe that our conclusions are robust. 
Indeed, the classical thermal bomb models have been very successful in explaining the basic features of the SN nucleosynthesis and the Galactic chemical evolution (\S \ref{ sec:intro } and references therein). 
The success of the new generation `1D calibrated' neutrino-driven models can be naturally understood as the outcome of the rapid explosion, in terms of the explosion timescale. Therefore, requiring the rapid explosion is a simple and straightforward solution for the SN mechanism to satisfy all these constraints, without fine tuning. 
Rather, it will be extremely fine-tuning if one would try to satisfy all the observational constraints in the context of the slow explosion.

Our results are summarized in Figure \ref{ fig:summary }. 
Figure \ref{ fig:summary } illustrates various observational constraints on the energy-time plane. 
The gray dotted and dashed lines show the time evolution of the explosion energy by \cite{2016ApJ...821...38S} (`artificial' explosion) and that by \cite{2016MNRAS.461L.112T} (ab-initio explosion), respectively. 
The hatched region shows the region where the production of the Fe-group elements is suppressed as compared to the instantaneous explosion, following an analytic treatment by  \cite{2009MNRAS.394.1317M}. 
It is seen that the region in the evolution of the explosion energy we reject from the present study is largely overlapping with their estimate. 
Figure \ref{ fig:summary } shows that energy growth must be sufficiently rapid so that its history passes above this hatched region. 
Otherwise, the peak temperature will already fall below $5\times10^9$ K when reaching to this region, and subsequent energy growth will not affect the iron group element synthesis.  
Therefore, Figure \ref{ fig:summary } shows that the nucleosynthesis yields are characterized by the energy generation in the first $\sim$100 msec after the initiation of the explosion. 
It is thus naturally understood that the observational success of the new generation `1D calibrated' neutrino-driven models is the outcome of the rapid explosion.
On the other hand, the slow mechanism ($t_{\rm grow} = 1000$ msec), which is suggested by the recent ab-initio simulations, passes through the region in the evolution of the energy which does not satisfy the observational constraints.

In summary, various observational constraints are consistent with the `rapid' explosion ($t_{\rm grow} \leqq 250$ msec), and especially the best match is found for a nearly instantaneous explosion ($t_{\rm grow} \lesssim 50$ msec). 
Our finding places a strong constraint on the explosion mechanism; the slow mechanism ($t_{\rm grow} \gtrsim 1000$ msec), which is suggested by recent ab-initio simulations, is rejected from these constraints (Figure \ref{ fig:summary }). 
We emphasize that we are not arguing against the standard framework of the delayed neutrino explosion scenario. 
We rather suggest that there must be a key ingredient still missing in the ab-initio simulations, which should lead to the rapid explosion.

\acknowledgments
The authors thank Yudai Suwa for stimulating discussion, and Ryoma Ouchi for kindly providing the progenitor model. 
The work has been supported by Japan Society for the Promotion of Science (JSPS) KAKENHI Grant 17H02864, 18H04585, 18H05223 (K.M.) and 19J14179 (R.S.). 

\software{blcode (Morozova et al. 2015) , MESA (v8118; Paxton et al. 2011, 2013, 2015)}

\bibliographystyle{aasjournal}

\bibliography{RS2019}

\end{document}